\documentclass[prd,reprint,nofootinbib,superscriptaddress,preprintnumbers,showpacs]{revtex4-1}
\usepackage{amsfonts}
\usepackage{amssymb}
\usepackage{amsmath}
\usepackage{graphicx,color}
\usepackage{dcolumn}
\usepackage{epsfig}
\usepackage{bm}
\usepackage{bbm}
\usepackage{ulem}
\usepackage{hyperref}
\usepackage{lineno}

\usepackage{color}

\newcommand{\ket}[1]{|{#1}\rangle}
\newcommand{\bra}[1]{\langle{#1}|}

\def\gtap{\ \raise.3ex\hbox{$>$\kern-.75em\lower1ex\hbox{$\sim$}}\ }
\def\ltap{\ \raise.3ex\hbox{$<$\kern-.75em\lower1ex\hbox{$\sim$}}\ }
\begin{document}

\title{
$P_c(4312)^+$, $P_c(4380)^+$, and $P_c(4457)^+$ as double triangle cusps
}
\author{Satoshi X. Nakamura}
\email{satoshi@ustc.edu.cn}
\affiliation{
University of Science and Technology of China, Hefei 230026, 
People's Republic of China
}
\affiliation{
State Key Laboratory of Particle Detection and Electronics (IHEP-USTC), Hefei 230036, People's Republic of China}

\begin{abstract}
Understanding the nature of 
the hidden charm pentaquark(like) signals in the LHCb data 
for $\Lambda_b^0\to J/\psi p K^-$
is a central problem of hadron spectroscopy.
We propose a scenario completely different from previous ones such as
 hadron molecules and compact pentaquarks. 
We identify relevant double triangle mechanisms
with leading or lower-order singularities.
The associated anomalous threshold cusps 
at the $\Sigma_c^{(*)}\bar{D}^{(*)}$ thresholds 
are significantly more singular than the ordinary ones.
Then we demonstrate that the double triangle amplitudes reproduce
the peak structures of $P_c(4312)^+$, $P_c(4380)^+$, and $P_c(4457)^+$
in the LHCb data, 
through an interference with other common mechanisms.
Only the $P_c(4440)^+$ peak is due to a resonance
with width and strength significantly smaller than previously
estimated. 
$P_c^+$ signals are expected
in other processes and
the proposed model (partly) explains the current data such as:
the GlueX $J/\psi$ photoproduction data with no 
 $P_c^+$ signals;
the LHCb $\Lambda_b^0\to J/\psi p \pi^-$ data 
with a possible signal only from $P_c(4440)^+$.
The double triangle singularity is now a possible option to
interpret resonancelike structures near thresholds in general.
\end{abstract}

\maketitle

\section{introduction}

Hadron spectrum is a reflection of
nonperturbative aspects of QCD.
Exotic hadrons have structures different from
the conventional quark-antiquark and three-quark~\cite{qm},
and are expected to provide a clue to assess a different aspect of QCD.
Recent experimental discoveries of exotic hadron(like) signals are therefore exciting~\cite{review_chen,review_hosaka,review_lebed,review_esposito,review_ali,review_guo,review_olsen,review_Brambilla}.

The LHCb Collaboration recently 
observed in $\Lambda_b^0\to J/\psi pK^-$ three resonance(like)
structures~\cite{lhcb_pc}, interpreted as
pentaquark states called $P_c(4312)^+$, $P_c(4440)^+$, and $P_c(4457)^+$,
by updating their
previous analysis~\cite{lhcb_pc_old}.
Understanding the nature of the $P_c^+$'s is certainly a central
problem of the hadron spectroscopy. 
Because of the proximity of the 
$P_c^+$ masses to the $\Sigma_c(2455)\bar{D}^{(*)}$ thresholds~\footnote{
We follow the hadron naming scheme of~\cite{pdg}. 
For simplicity, however, we often
denote $\Sigma_c(2455)^{+(++)}$, $\Sigma_c(2520)^{+(++)}$,
$\Lambda_c(2595)^+$, $\Lambda_c(2625)^+$, and $J/\psi$ 
by $\Sigma_c$, $\Sigma_c^*$, $\Lambda^*_c$, $\Lambda^{**}_c$, 
and $\psi$, 
respectively.
$\Lambda^{(*,**)}_c$ and $\Sigma_c^{(*)}$ are also collectively denoted by 
$Y_c$.
Charge indices are often suppressed.
},
$P_c^+$'s as $\Sigma_c(2455)\bar{D}^{(*)}$ molecules (bound states) may seem a natural
interpretation of their identity~\cite{pc_beihang,pc_valencia,pc_beihang2,pc_hebei,pc_nanjin,pc_itp,pc_chen,pc_lanzhou,pc_wang,pc_gutsche,pc_peking,pc_nanjing2,pc_lin,pc_burns,pc_xu,pc_yamaguchi,pc_sakai,pc_voloshin,pc_wu,pc_jrzhang,pc_hxu,pc_du,pc_du2,pc_xiao}.
Yet, constituent pentaquark
pictures~\cite{pc_ali,pc_pimikov,pc_zgwang,pc_rzhu,pc_xzweng,pc_bari,pc_stancu,pc_ydong} 
and a hadrocharmonium~\cite{pc_hadrochamonium}
are also possible options.
The LHCb data of the $J/\psi p$ invariant mass ($M_{J/\psi p}$)
distribution has been analyzed with 
a $K$-matrix model 
that claimed a virtual state for $P_c(4312)^+$~\cite{jpac}.
Another analyses~\cite{pc_du,pc_du2,pc_xiao}
based on one-pion-exchange plus contact interactions
for the coupled $\Sigma_c^{(*)}\bar{D}^{(*)}$ system 
interpreted all the $P_c^+$'s as $\Sigma_c^{(*)}\bar{D}^{(*)}$ bound states;
they also claimed the existence of a narrow $P_c(4380)^+$~\footnote{
In this paper, 
$P_c(4380)^+$ does not refer to
a broad state~\cite{lhcb_pc_old}.
}.
So far, all the models assigned a pole
to each of the $P_c^+$ peaks~\footnote{
Although a $K$-matrix analysis~\cite{pc_kuang} claimed $P_c(4457)^+$ as a threshold
cusp, we put the conclusion on hold 
for insufficient quality of fitting the $P_c(4312)^+$ peak and 
$\Sigma_c(2520)\bar{D}$ threshold region.
}.

In order to establish $P_c^+$'s as hadronic states, 
it is important to confirm their signals in other processes such as 
$J/\psi$ photoproduction off a nucleon~\cite{photo_qwang,photo_kubarovsky,photo_Karliner,photo_hiller,photo_xywang,photo_wu,photo_cao}.
The GlueX Collaboration conducted such an experiment,
finding no evidence~\cite{gluex}.
This may indicate that
the $P_c^+$ states couple weakly with a photon and could be seen in
higher statistics data.
Another possibility is that
the $P_c^+$ peaks in $\Lambda_b^0\to J/\psi pK^-$ are caused by
kinematical effects such as 
threshold cusp and triangle singularity (TS)~\cite{ts_review},
and do not appear in the photoproduction;
however, no relevant mechanism has been found.
TS~\cite{ts1,ts2} proposed for
the previous $P_c^+$~\cite{lhcb_pc_old}
no longer fit the updated $P_c^+$ signals unless a pole is included~\cite{lhcb_pc}.

\begin{figure*}
\begin{center}
\includegraphics[width=1\textwidth]{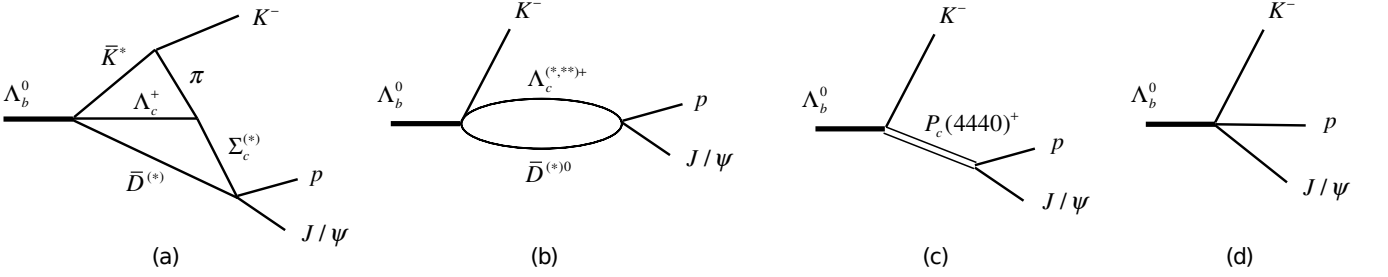}
\end{center}
 \caption{
$\Lambda_b^0\to J/\psi pK^-$ mechanisms:
(a) double triangle;
(b) one-loop;
(c) $P_c(4440)^+$-excitation;
(d) direct decay.
 }
\label{fig:diag}
\end{figure*}

In this work, we point out that 
a double triangle (DT) diagram [Fig.~\ref{fig:diag}(a)]
creates a kinematical effect
that has not been explored for interpreting resonancelike structures.
The effect is caused by the fact that
each loop hits a TS
and that
the two TS can even occur almost simultaneously (leading singularity~\cite{s-matrix}).
The associated anomalous threshold cusps are 
significantly more singular than the ordinary ones
and, thus, can be a new option to understand
exotic hadronlike signals near thresholds.
We then demonstrate that the DT amplitudes reproduce
the peak structures of $P_c(4312)^+$, $P_c(4380)^+$, and $P_c(4457)^+$
in the LHCb data, 
through an interference with other common mechanisms
[Figs.~\ref{fig:diag}(b) and \ref{fig:diag}(d)].
The LHCb data requires this proposed model to have
only $P_c(4440)^+$ as a resonance
with width and strength significantly smaller than previously
estimated. 
The model can also (partly) explain $P_c^+$ signals in other data such as:
no $P_c^+$ signals in the $J/\psi$ photoproduction;
$\Lambda_b^0\to J/\psi p \pi^-$ data~\cite{Pc_lhcb2}
suggesting only a $P_c(4440)^+$ signal.
In this way, we bring a completely new understanding of the $P_c^+$
structures in the LHCb data.

\section{model}
Our model for $\Lambda_b^0\to J/\psi pK^-$
is diagrammatically represented in 
Fig.~\ref{fig:diag}.
For loop diagrams of Fig.~\ref{fig:diag}(a,b), 
we assume that color-favored
$\Lambda_b^0\to \Lambda^{(*,**)+}_c\bar{D}^{(*)}\bar{K}^{(*)}$
decays dominate over 
color-suppressed ones.
We do not include color-suppressed 
$\Lambda_b^0\to \Sigma^{(*)}_c\bar{D}^{(*)}K^-$ vertices
that previous models often used 
with a possible problem of explaining $P_c^+$ production
rates~\cite{pc_burns}.
We could include color-suppressed processes since
their suppression is generally difficult to predict~\cite{pc_du2} and
the DT amplitudes might be
more suppressed.
We still take this assumption because
the DT amplitudes are actually not significantly suppressed
and, for fitting only the $M_{J/\psi p}$ distribution data,
color-suppressed contributions are 
redundant and would not significantly improve the fit.
Regarding the parity,
we consider only parity-conserving mechanisms;
parity-violating mechanisms should exist but are redundant in the fits.
Partial waves of
$J^P=1/2^-$, $3/2^-$, $1/2^+$, and $3/2^+$ are considered;
$J^P$ denotes the spin-parity of $J/\psi p$.

We present amplitude
formulas for representative diagrams in Fig.~\ref{fig:diag};
see the Supplemental material for
complete formulas and parameter values.
We use the particle mass and width values from \cite{pdg},
and denote the energy, width, momentum, and polarization vector of a particle $x$ by
$E_x$, $\Gamma_x$, $\bm{p}_x$, and $\bm{\epsilon}_x$, respectively.
We also denote a baryon($B$)-meson($M$) pair by $BM(J^P)$.
A DT diagram
[Fig.~\ref{fig:diag}(a)]
that includes $\Sigma_c\bar{D}\, (1/2^-)$
consists of 
four vertices such as 
$\Lambda_b^0\to \Lambda_c^+\bar{D}\bar{K}^{*}$, 
$\bar{K}^*\to \bar{K}\pi$,
$\Lambda_c^+\pi\to\Sigma_c$, and 
$\Sigma_c\bar{D}\,(1/2^-)\to J/\psi p$ given by
\begin{eqnarray}
\label{eq:lps}
&& c_{\Lambda_c\bar{D}\bar{K}^{*},\Lambda_b}
\bigg({1\over 2} t_{\bar{D}} {1\over 2} t_{\bar{K}^*}\bigg|00\bigg)\,
\bm{\sigma}\cdot\bm{\epsilon}_{\bar{K}^*}\,
 F_{\bar{K}^*\bar{D}\Lambda_c,\Lambda_b}^{00}\ , \\
&&c_{\bar{K}\pi,\bar{K}^*}\, 
\bigg(1 t_\pi {1\over 2} t_{\bar{K}}\bigg| {1\over 2} t_{\bar{K}^*}\bigg)\,
\bm{\epsilon}_{\bar{K}^*} \cdot ( \bm{p}_{\bar{K}}-\bm{p}_{\pi})\,
 f_{\bar{K}\pi,\bar{K}^*}^{1}
\ , \\ 
&&c_{\Lambda_c\pi,\Sigma_c}\, 
\bm{\sigma} \cdot \bm{p}_{\pi} \,
 f_{\Lambda_c\pi,\Sigma_c}^{1}
\ , \\
&& c^{1/2^{-}}_{\psi p, \Sigma_c\bar{D}}\,
\bigg(1 t_{\Sigma_c} {1\over 2} t_{\bar{D}}\bigg| {1\over 2} t_p\bigg)\,
\bm{\sigma}\cdot \bm{\epsilon}_\psi \,
 f_{\psi p}^{0}\,
 f_{\Sigma_c\bar{D}}^{0} \ ,
\end{eqnarray}
respectively.
Dipole form factors with a cutoff $\Lambda$
are denoted by
$F_{ijk,l}^{LL'}$,
$f_{ij}^{L}$, and $f_{ij,k}^{L}$.
The parentheses are isospin Clebsch-Gordan coefficients (CGCs) with 
$t_x$ being the isospin $z$-component of $x$.
In Eq.~(\ref{eq:lps}) and later in Eq.~(\ref{eq:lps2}),
the isospin of the $\bar{D}^{(*)}\bar{K}^{(*)}$ pair is 0
for assuming the color-favored $\Lambda_b^0$ decays.
The couplings
$c_{\bar{K}\pi,\bar{K}^*}$ and $c_{\Lambda_c\pi,\Sigma_c}$
are determined by the $K^*$ and $\Sigma_c(2455)$ decay widths.
The DT amplitude is
\begin{eqnarray}
 A^{\rm DT}_{\Sigma_c\bar{D}(1/2^-)} &=&
\sum_{\rm charge}
c^{1/2^{-}}_{\psi p, \Sigma_c\bar{D}}\,
c_{\Lambda_c\pi,\Sigma_c}\, 
c_{\bar{K}\pi,\bar{K}^*}\, 
c_{\Lambda_c\bar{D}\bar{K}^{*},\Lambda_b}
\nonumber\\
&&\times
({\rm isospin\ CGCs})
\int d^3p_{\pi}
\int d^3p_{\bar{D}}
\nonumber\\
&&\times
{\bm{\sigma}\cdot \bm{\epsilon}_\psi\,
\bm{\sigma}\cdot \bm{p}_\pi \,
\over E-E_{\bar{K}}-E_{\Sigma_c}-E_{\bar{D}}+{i\over 2}\Gamma_{\Sigma_c}}
\nonumber\\
&&\times
{\bm{\sigma}\cdot (\bm{p}_{\bar{K}}-\bm{p}_\pi)
\over E-E_{\bar{K}}-E_\pi-E_{\Lambda_c}-E_{\bar{D}}+i\epsilon}\nonumber \\
&&\times
{ f_{\psi p}^{0}\,
 f_{\Sigma_c\bar{D}}^{0}
 f_{\Lambda_c\pi,\Sigma_c}^{1}
 f_{\bar{K}\pi,\bar{K}^*}^{1}
 F_{\bar{K}^*\bar{D}\Lambda_c,\Lambda_b}^{00}
\over E-E_{\bar{K}^*}-E_{\Lambda_c}-E_{\bar{D}}+{i\over 2}\Gamma_{K^*}} \ ,
\label{eq:DT}
\end{eqnarray}
where 
different charge states in the loops are summed with
charge dependent particle masses; $E$ is the total energy.
The amplitude is implicitly sandwiched by 
initial $\Lambda_b^0$ and final $p$ spin states.

The $\Lambda_c^+\bar{D}^{*0}\,(1/2^-)$ one-loop amplitude [Fig.~\ref{fig:diag}(b)]
is composed by an initial 
$\Lambda_b^0\to \Lambda_c^+\bar{D}^{*0}K^-$
vertex
\begin{eqnarray}
 c_{\Lambda_c\bar{D}^*\bar{K},\Lambda_b}
\bigg({1\over 2} t_{\bar{D}^*} {1\over 2} t_{\bar{K}}\bigg|00\bigg)\,
\bm{\sigma}\cdot\bm{\epsilon}_{\bar{D}^*}
 F_{\bar{K}\bar{D}^*\Lambda_c,\Lambda_b}^{00}\ , \
\label{eq:lps2}
\end{eqnarray}
and the subsequent
$\Lambda_c^+\bar{D}^{*0}\,(1/2^-)\to J/\psi p$
interaction
\begin{eqnarray}
c^{1/2^-}_{\psi p, \Lambda_c\bar{D}^*}
\bm{\sigma}\cdot \bm{\epsilon}_\psi \,
\bm{\sigma}\cdot\bm{\epsilon}_{\bar{D}^*}
 f_{\psi p}^{0}
 f_{\Lambda_c\bar{D}^*}^{0} \ .
\end{eqnarray}
The one-loop amplitude is
\begin{eqnarray}
 A^{\rm 1L}_{\Lambda_c\bar{D}^*} &=&
3 \, 
 c^{1/2^{-}}_{\psi p, \Lambda_c\bar{D}^*}\,
 c_{\Lambda_c\bar{D}^*\bar{K},\Lambda_b}\,
\bigg({1\over 2} t_{\bar{D}^*} {1\over 2} t_{\bar{K}}\bigg|00\bigg)\,
\bm{\sigma}\cdot \bm{\epsilon}_\psi 
\nonumber\\
&&\times\!
\int d^3p_{\bar{D}^*} 
 {
 f_{\psi p}^{0}
 f_{\Lambda_c\bar{D}^*}^{0}
 F_{\bar{K}\bar{D}^*\Lambda_c,\Lambda_b}^{00}
\over E-E_{\bar{K}}-E_{\Lambda_c}-E_{\bar{D}^*}+{i\over 2}\Gamma_{D^*}} ,
\label{eq:1L}
\end{eqnarray}
with $\Gamma_{D^{*0}}=55$~keV~\cite{sxn_x}.

The $P_c(4440)^+$ amplitude [Fig.~\ref{fig:diag}(c)] is given 
in the Breit-Wigner form
with adjustable
$P_c(4440)^+$ mass, width, and
coupling.
For $J^P$ of $P_c(4440)^+$,
we examine $1/2^\pm$ and $3/2^\pm$ cases.
The internal structure of $P_c(4440)^+$
is beyond the scope of this work.
We also consider 
a direct decay mechanism [Fig.~\ref{fig:diag}(d)]
in each partial wave with a real coupling strength. 
This simulates all mechanisms
not belonging to the diagrams of 
Figs.~\ref{fig:diag}(a)-\ref{fig:diag}(c).

The $Y_c\bar D^{(*)}$ pairs in the
DT and one-loop diagrams
require a nonperturbative treatment. 
Meanwhile, the lattice QCD~\cite{lqcd_jpsi_p} found
a weak $J/\psi p\to J/\psi p$ interaction.
Thus, a reasonable approach is to develop a 
$Y_c\bar D^{(*)}$ coupled-channel model incorporating the heavy quark
spin symmetry~\cite{pc_du,pc_du2,pc_xiao,pc_yamaguchi}.
However, we take a simpler approach and 
use a single-channel contact interaction model~\cite{d-decay} to calculate 
a $Y_c\bar D^{(*)}$
scattering amplitude that maintains only
the elastic unitarity.
Then, a perturbative transition to $J/\psi p$ follows.
The resulting 
$Y_c\bar D^{(*)}\to J/\psi p$ amplitude 
is implemented in Eqs.~(\ref{eq:DT}) and (\ref{eq:1L}); see the
Supplemental material.
Other possible coupled-channel effects are assumed to be absorbed by complex
couplings fitted to the data. 
This simplification can be justified
for the limited purpose of proposing a new scenario for 
the $P_c^+$ structures in the LHCb data.
Within our model, 
the $P_c^+$ structures (except for $P_c(4440)^+$) are
mostly described by the kinematical
effects, and not directly by poles from the $Y_c\bar D^{(*)}$ scattering.
Therefore, the data can only loosely constrain
our $Y_c\bar D^{(*)}$ interactions.
Conversely, the details of 
the $Y_c\bar D^{(*)}$ scattering do not play a major role.
This is in contrast with hadron-molecule models for which 
the $Y_c\bar D^{(*)}$ interactions must be fine-tuned to get poles at
right positions; the details do matter.

\section{results}

\subsection{Double triangle amplitudes and their singular behaviors}

\begin{figure}[t]
\begin{center}
\includegraphics[width=.5\textwidth]{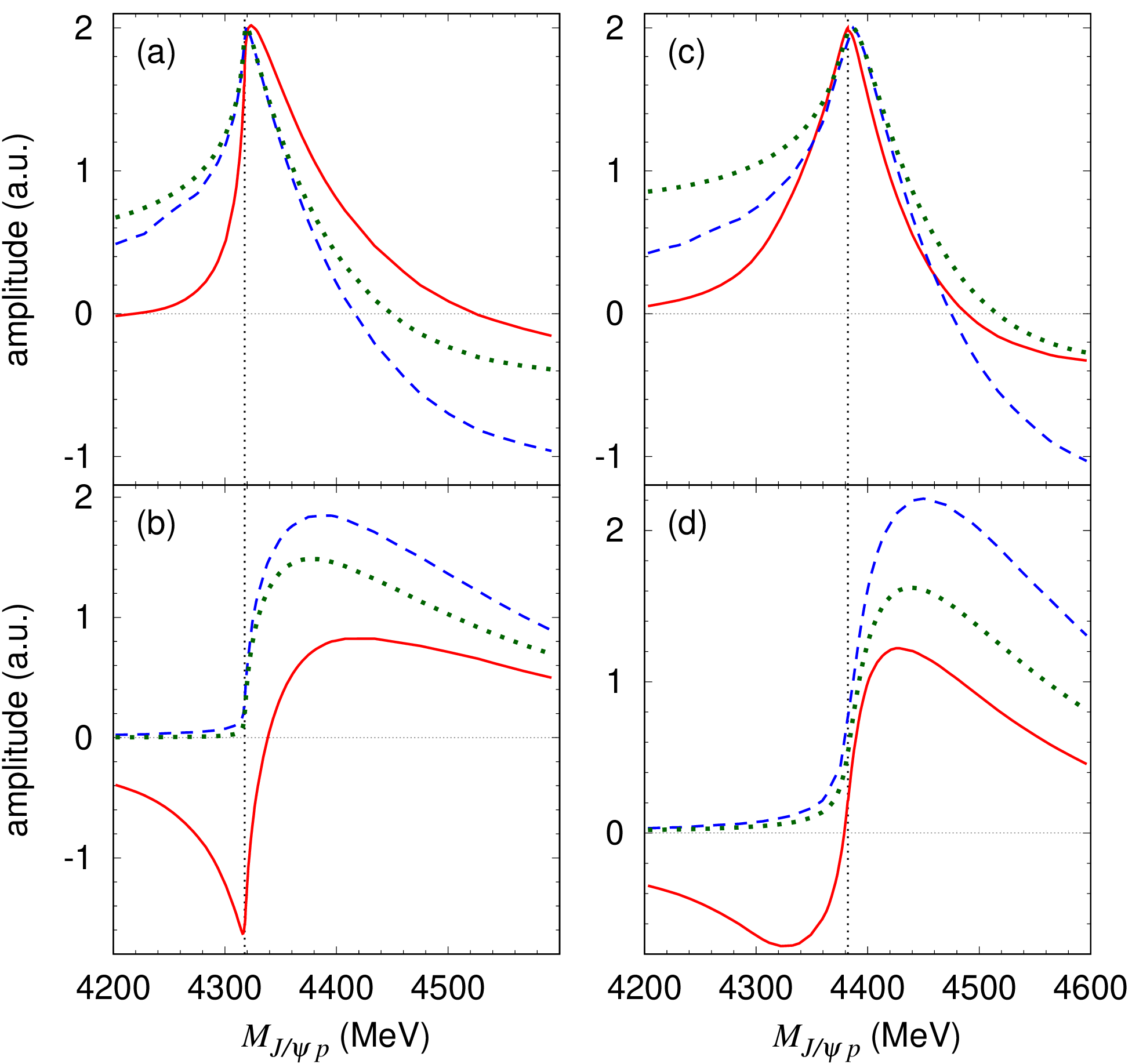}
\end{center}
 \caption{
Double triangle amplitudes.
The panel (a) [(b)] shows the real [imaginary] part.
The red solid curves are for
$\Sigma_c^{(*)}\bar D^{(*)}=\Sigma_c^{+}\bar D^{0}$
in Fig.~\ref{fig:diag}(a) [$\Lambda=1$~GeV;
perturbative $\Sigma^{(*)}_c\bar{D}^{(*)}\to J/\psi p$].
They reduce to the blue dashed ones
by using $m_{\Lambda_c^+}=3$~GeV.
The green dotted curves are the
$\Sigma_c^{+}\bar D^{0}$ one-loop amplitude.
All the amplitudes are normalized so that the real parts 
have the same peak height. 
The dotted vertical lines indicate
the $\Sigma_c^{+}\bar D^{0}$ thresholds.
The amplitudes shown in the panel (c) [(d)] are obtained from those in
(a) [(b)] by replacing
$\Sigma_c^{+}$ with
 $\Sigma_c^{*+}$.
 }
\label{fig:amp}
\end{figure}
The DT amplitudes cause
the leading and lower-order singularities (or anomalous thresholds)~\cite{s-matrix} 
when $\Sigma_c$ and $\Sigma^*_c$ propagates in Fig.~\ref{fig:diag}(a), respectively.
This is analogous to a single triangle diagram causing TS as
its leading singularity.
According to the Coleman-Norton theorem~\cite{coleman},
the leading singularity occurs 
when a given loop process is kinematically allowed at the
classical level.
The lower-order singularity occurs
when one (or more) of the intermediate states is necessarily off-shell.
Details on the kinematical conditions 
are given
in the Supplemental material. 
As a TS-enhanced amplitude shows 
a logarithmically singular behavior, 
the DT leading and lower-order singularities 
generate their own singular behaviors. 
While the singular behaviors can, in principle, be examined by analyzing the corresponding
Landau equations~\cite{s-matrix,landau},
we study them numerically below.

We plot in Fig.~\ref{fig:amp}(a,b)
the DT amplitude that
includes $\Sigma_c^{+}\bar{D}^{0}\,(1/2^-)$ [red solid curves].
A singular behavior clearly shows up near the 
$\Sigma_c^{+}\bar{D}^{0}$ threshold.
We also plot a threshold cusp,
due to the square-root singularity, 
from a $\Sigma_c^{+}\bar{D}^{0}$
one-loop amplitude similar to Fig.~\ref{fig:diag}(b)
[green dotted curves];
the imaginary part does not vanish below the threshold for
$\Gamma_{\Sigma_c^+}\ne 0$.
The DT cusp due to the leading singularity
is clearly more singular.
For an illustration,
we replace the $\Lambda_c^+$ mass in the DT amplitude with 
a hypothetically heavy value (3~GeV)
so that the loop integral can hit only the
$\Sigma_c^{+}\bar{D}^{0}$ threshold singularity.
The resultant amplitude [blue dashed curves]
behaves like an ordinary threshold cusp.
The amplitudes shown in Fig.~\ref{fig:amp}(a,b)
include $\Sigma^{+}_c\bar{D}^{0}\to J/\psi p$ 
treated as the first-order perturbation.
A non-perturbative attractive 
$\Sigma^{+}_c\bar{D}^{0}$ interaction significantly enhances
the singular behaviors~\cite{xkdong}.
Similar plots in Fig.~\ref{fig:amp}(c,d) 
are obtained 
by replacing $\Sigma^{+}_c$ in the amplitudes shown in 
Fig.~\ref{fig:amp}(a,b) with $\Sigma^{*+}_c$.
While this DT amplitude with $\Sigma^{*+}_c\bar{D}^0$ has 
the lower-order singularity and
is more singular than the ordinary threshold cusp,
it is less singular than the DT amplitude with $\Sigma^{+}_c\bar{D}^0$
having the leading singularity.

\begin{figure}[t]
\begin{center}
\includegraphics[width=.5\textwidth]{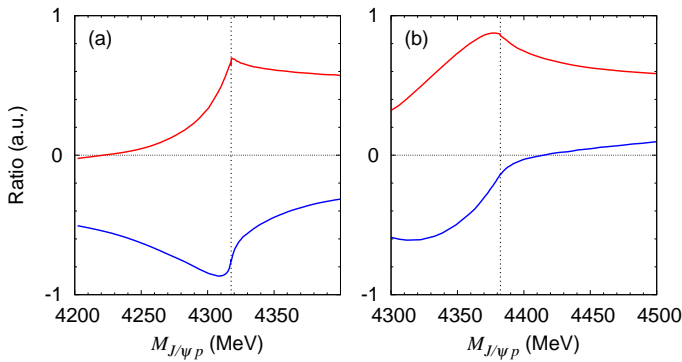}
\end{center}
 \caption{Ratio of double triangle ($A_{\rm DT}$) to one-loop
 ($A_{\rm 1L}$) amplitudes.
The ratio in the panel (a) [(b)] is from 
the amplitudes including $\Sigma^{+}_c\bar{D}^0$ [$\Sigma^{*+}_c\bar{D}^0$]
intermediate states.
The upper red (lower blue) curves show
${\rm Re}[A_{\rm DT}/A_{\rm 1L}]$ (${\rm Im}[A_{\rm DT}/A_{\rm 1L}]$).
The dotted vertical lines in the panel (a) [(b)]
indicate the $\Sigma_c^{+}\bar D^{0}$ [$\Sigma_c^{*+}\bar D^{0}$]
threshold.
 }
\label{fig:amp2}
\end{figure}
To clarify that the DT amplitudes and
ordinary threshold cusps have different 
singular behaviors, we divide the former by the latter and show the
ratios in Fig.~\ref{fig:amp2}.
The singular behaviors still remain in 
the real part of the ratios.
The first [second] derivative of the ratio in Fig.~\ref{fig:amp2}(a) [\ref{fig:amp2}(b)]
seems divergent at the 
$\Sigma_c^{+}\bar D^{0}$ [$\Sigma_c^{*+}\bar D^{0}$] threshold,
showing a qualitative difference between the leading and lower-order
singularities.
The results in Figs.~\ref{fig:amp}-\ref{fig:amp2}
indicate that, in general,
the DT cusp might be
an option to understand
resonancelike structures near thresholds.

As expected from Fig.~\ref{fig:amp},
the DT amplitude alone creates a peak at 
the $\Sigma_c^{(*)}\bar{D}^{(*)}$ threshold
in the $M_{J/\psi p}$ distribution of
$\Lambda_b^0\to J/\psi pK^-$.
This peak cannot be identified with a $P_c^+$ peak which is located
slightly below the threshold.
However, 
suppose there exists a smooth amplitude that
interferes with the real (imaginary) part of 
the DT amplitude of Fig.~\ref{fig:amp}
destructively (constructively) 
near the threshold.
This coherent sum generates a peak 
slightly below the $\Sigma_c^{(*)}\bar{D}^{(*)}$ threshold.
This is how the $P_c^+$-like structures (other than $P_c(4440)^+$)
show up from the DT mechanisms.

\subsection{Analysis of the LHCb data}

\begin{figure*}[t]
\begin{center}
\includegraphics[width=1\textwidth]{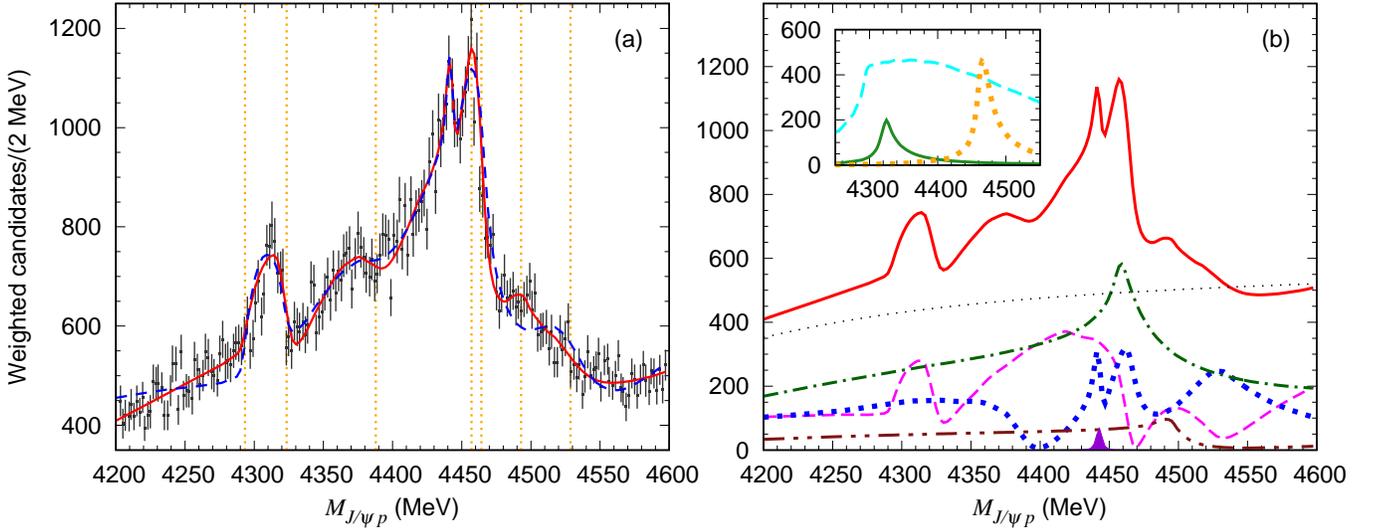}
\end{center}
 \caption{
$J/\psi p$ invariant mass $(M_{J/\psi p})$ distribution for
$\Lambda_b^0\to J/\psi pK^-$.
(a) Comparison with the LHCb data
($\cos\theta_{P_c}$-weighted samples)~\cite{lhcb_pc}.
The red solid (blue dashed)
curve is from the full (simplified) model;
see the text for details on the models.
The dotted vertical lines indicate thresholds for,
 from left to right, $\Lambda_c^+\bar{D}^{*0}$, $\Sigma_c(2455)^{++}D^-$,
$\Sigma_c(2520)^{++}D^-$, 
$\Lambda_c(2595)^+\bar{D}^0$,
$\Sigma_c(2455)^{++}D^{*-}$, 
$\Lambda_c(2625)^+\bar{D}^0$,
and $\Sigma_c(2520)^{++}D^{*-}$, respectively.
(b)~Partial wave contributions
specified by $J^P$ of the $J/\psi p$ pair.
The magenta dashed, blue dotted, 
green dash-dotted, and brown dash-two-dotted
curves are for
$J^P=1/2^-, 3/2^-, 1/2^+$, and $3/2^+$, respectively.
The solid violet peak is the $P_c(4440)^+(3/2^-)$ contribution without interference.
The thin black dotted curve is the sum of the direct decay mechanisms. 
The red solid curve is from the full model.
In the insert, the green solid, orange dotted, and cyan dashed curves
 are contributions from 
the double triangle mechanism with
$\Sigma_c\bar{D}\, (1/2^-)$,
that with
$\Sigma_c\bar{D}^*\, (1/2^-)$,
and the $\Lambda_c^+\bar{D}^{*0}\, (1/2^-)$ one-loop mechanism, respectively, without the interference.
 }
\label{fig:comp-data}
\end{figure*}
For describing $\Lambda_b^0\to J/\psi pK^-$, 
our full model includes:
(i) DT mechanisms with
$\Sigma_c(2455)\bar{D}\, (1/2^-)$,
$\Sigma_c(2520)\bar{D}\, (3/2^-)$,
$\Sigma_c(2455)\bar{D}^*\, (1/2^-)$,
$\Sigma_c(2455)\bar{D}^*$ $(3/2^-)$,
$\Sigma_c(2520)\bar{D}^*\, (1/2^-)$, and
$\Sigma_c(2520)\bar{D}^*\, (3/2^-)$;
(ii) one-loop mechanisms with
$\Lambda_c^+\bar{D}^{*0}\, (1/2^-)$,
$\Lambda_c(2595)^+\bar{D}^{0}\, (1/2^+)$, and
$\Lambda_c(2625)^+\bar{D}^{0}\, (3/2^+)$;
(iii) $P_c(4440)^+$ mechanism;
(iv) direct decay mechanisms.
Regarding the number of fitting parameters, 
each mechanism in the items (i)-(iii) has an adjustable complex overall factor
to fit the data; $2\times 10$ parameters.
Each direct decay mechanism (iv) has a real coupling strength; 4 parameters.
The $P_c(4440)^+$ mass and width are adjustable.
We also adjust a repulsive $\Lambda_c^+\bar{D}^{*0}$ interaction strength.
Because the overall absolute normalization of the full amplitude is arbitrary, 
we have totally 26 parameters. 
Although the heavy quark spin symmetry
may constrain the parameters, we adjust them rather freely in the fits. 
This can be justified because 
the fitting parameters can effectively absorb effects from mechanisms 
not explicitly considered such as parity-violating amplitudes;
more discussions in the Supplemental material.

Regarding the elastic $Y_c\bar D^{(*)}$ 
interaction strengths, 
we examine if the fit favors
an attractive or repulsive 
interaction for a given $BM$ $(J^P)$.
The fit favors attractions for
$\Sigma_c(2455)\bar{D}\, (1/2^-)$,
$\Sigma_c(2520)\bar{D}\, (3/2^-)$,
$\Sigma_c(2455)\bar{D}^*\, (1/2^-)$,
$\Sigma_c(2455)\bar{D}^*\, (3/2^-)$,
$\Lambda_c(2595)^+\bar{D}^{0}$ $(1/2^+)$,
$\Lambda_c(2625)^+\bar{D}^{0}$ $(3/2^+)$,
and repulsions for
$\Sigma_c(2520)\bar{D}^*\, (1/2^-)$,
$\Sigma_c(2520)\bar{D}^*\, (3/2^-)$,
$\Lambda_c^+\bar{D}^{*0}\, (1/2^-)$.
Then we  fix the coupling strength 
so that the scattering length is $a \sim 0.5$~fm~\footnote{
The scattering length $(a)$ is related to the phase shift $(\delta)$ by 
$p\cot\delta=1/a + {\cal O}(p^2)$.} 
for the attraction.
The repulsive $\Lambda_c^+\bar{D}^{*0}\,(1/2^-)$ interaction strength
is fitted to the data because the fit quality is rather sensitive;
$a \sim -0.4$, $-0.2$, and $-0.05$~fm
for $\Lambda\sim 0.8$, $1$, and $1.5$--$2$~GeV, respectively.
The other repulsive interactions have the same strength as 
$\Lambda_c^+\bar{D}^{*0}$.
It is noted that, within our model,
spectrum peak positions are not very sensitive to the $a$ values.
The cutoff in the form factors is fixed at $\Lambda=1$~GeV unless
otherwise stated.
Only the direct decay amplitudes include
different cutoffs on $p_{\bar{K}}$
so that their $M_{J/\psi p}$ distribution is similar to the phase-space shape.

We compare
our calculation with the LHCb data~\cite{lhcb_pc} in Fig.~\ref{fig:comp-data}(a).
The experimental resolution is considered in the calculation. 
Our full model (red solid curve) well fits the data.
The $P_c(4312)^+$, $P_c(4380)^+$, and $P_c(4457)^+$ 
peaks are well described by the kinematical effects from the
considered mechanisms, and not by poles near the 
peak positions.
Only the $P_c(4440)^+$ peak requires a resonance pole for which we
choose $J^P=3/2^-$ in the figure.
The fit quality does not significantly change when
varying the cutoff over $\Lambda=$ 0.8--2~GeV
and when choosing $J^P=1/2^\pm$ and $3/2^\pm$ for $P_c(4440)^+$;
the fits thus do not favor a particular $J^P$ of $P_c(4440)^+$.
We simplify the model by omitting 
the $1/2^+$ and $3/2^+$ amplitudes, and treating
$Y_c\bar{D}^{(*)}\to J/\psi p$
perturbatively.
With 19 parameters to refit,
main features of the data are fairly well captured (blue dashed curve).
Still, the structures are less sharp 
near the thresholds because of lacking the
strong $Y_c\bar{D}^{(*)}$ scattering.
The $1/2^+$ and $3/2^+$ amplitudes are also needed 
for a more precise fit.

In Fig.~\ref{fig:comp-data}(b),
we show 
the full model's
partial wave contributions that do not interfere with each other 
in the $M_{J/\psi p}$ distribution.
In the $1/2^-$ (magenta dashed curve) and 
$3/2^-$ (blue dotted curve) contributions,
the interference between the DT and direct
decay mechanisms creates bumps slightly below 
the $\Sigma_c^{(*)}\bar{D}^{(*)}$ thresholds
where the $P_c(4312)^+$, $P_c(4380)^+$, and $P_c(4457)^+$ peaks are located.
The $1/2^-$ contribution also shows
a bent due to the $\Lambda_c^+\bar{D}^{*0}$ threshold cusp,
being consistent with the data. 
The $1/2^+$ (green dash-dotted curve) and 
$3/2^+$ (brown dash-two-dotted curve) contributions
exhibit 
$\Lambda_c(2595)^+\bar{D}^0$ and $\Lambda_c(2625)^+\bar{D}^0$
threshold cusps, respectively, capturing the characteristic
structures in the data. 
Although the $1/2^+$ peak seems a large contribution, 
this is due to a constructive interference between 
the $\Lambda_c(2595)^+\bar{D}^0$ one-loop and direct decay amplitudes;
the $\Lambda_c(2595)^+\bar{D}^0$ one-loop amplitude itself is
significantly smaller than 
the $\Lambda_c^+\bar{D}^{*0}$ one-loop amplitude in magnitude.
The sum of the direct decay mechanisms 
shows a phase-space-like distribution (thin black dotted curve).
This partial wave decomposition would involve uncertainty due to
the limited experimental information.

The $P_c(4440)^+(3/2^-)$ contribution without interference
is shown in Fig.~\ref{fig:comp-data}(b) [violet solid peak].
The mass and width from our fit 
is 4443.1$\pm 1.4$~MeV and 2.7$\pm 2.4$~MeV,
respectively;
the statistical errors 
are estimated just for a reference
by varying 
only the $P_c(4440)^+$ mass, width, and coupling.
The cutoff dependence is safely within the errors.
Comparing with the LHCb analysis~\cite{lhcb_pc}, $4440.3\pm 1.3^{+4.1}_{-4.7}$~MeV and 
$20.6\pm 4.9^{+8.7}_{-10.1}$~MeV, 
the width is significantly narrower. 
Also, the $P_c(4440)^+$ contribution 
is only $\sim$1/22
of the LHCb's estimate:
${\cal R}\equiv {\cal B}(\Lambda_b^0\to P_c^+K^-){\cal B}(P_c^+\to J/\psi p)
/{\cal B}(\Lambda_b^0\to J/\psi pK^-)=1.11\pm 0.33^{+0.22}_{-0.10}$~\%.
This difference arises because
the LHCb fitted the structure at 
$M_{J/\psi p}\sim 4450$~MeV with 
incoherent $P_c(4440)^+$ and $P_c(4457)^+$ contributions
while we describe a large portion of the structure with the
kinematical effects and 
attribute only the small spike 
to the $P_c(4440)^+$ and its interference. 

The quality of the fit could be slightly improved with
more attractive $\Sigma_c^{(*)}\bar{D}^{(*)}$ interactions.
However, such interactions generate 
virtual poles near the thresholds and, as a result, 
$\gamma p\to J/\psi p$ cross section would have
sharp peaks at the $\Sigma_c^{(*)}\bar{D}^{(*)}$ thresholds.
Since the GlueX did not find such peaks~\cite{gluex}, 
we fix the $\Sigma_c^{(*)}\bar{D}^{(*)}$ interactions at the moderate
strength so that no poles are generated close to the thresholds~\footnote{
Our $\Sigma_c\bar{D}$ interaction model
with $a\sim 0.5$~fm
generates a virtual pole at $\sim$20~MeV below the threshold.
}.

To examine whether 
the DT amplitudes are significantly suppressed
compared to other mechanisms,
we show the DT and one-loop contributions to
the $1/2^-$ partial wave 
in the insert of Fig.~\ref{fig:comp-data}(b).
The peak height of the DT contributions is comparable to that of the
one-loop contribution. 
We can further compare the coupling strength of 
the DT amplitude [Eq.~(\ref{eq:DT})]
to that of the $\Lambda_c^+\bar{D}^{*0}$ one-loop amplitude
[Eq.~(\ref{eq:1L})] by 
introducing a ratio:
\begin{eqnarray}
R = 
\left|{c^{1/2^{-}}_{\psi p, \Sigma_c\bar{D}}\,
c_{\Lambda_c\bar{D}\bar{K}^{*},\Lambda_b} \over
c^{1/2^{-}}_{\psi p, \Lambda_c\bar{D}^*}\,
c_{\Lambda_c\bar{D}^*\bar{K},\Lambda_b} }\right| \ ,
\end{eqnarray}
where well-controlled parameters ($c_{\bar{K}\pi,\bar{K}^*}$, $c_{\Lambda_c\pi,\Sigma_c}$) 
are excluded.
Fitting the LHCb data gives $R= 7.2-3.2$ for $\Lambda = 0.8-2$~GeV. 
Thus, 
unreasonably large parameter values $(R\gg 1)$ are not necessary,
indicating that the comparable strengths of 
the DT and one-loop amplitudes are not an artifact.

Our model partly explains the absence of $P_c^+$ signals 
in $J/\psi$ photoproduction data~\cite{gluex}.
All $P_c^+$ peaks, except for $P_c(4440)^+$, in 
$\Lambda_b^0\to J/\psi pK^-$ are caused by the kinematical effect 
(DT mechanisms and its interference)
that requires an accompanying $K^-$
in the final state.
The $\gamma p\to J/\psi p$ process
cannot accommodate this kinematical effect 
and, thus, has no $P_c^+$ signals.
For $P_c(4440)^+$, 
its width and fit fraction from our analysis 
are significantly smaller than those from the LHCb's.
In this case, observing a $P_c(4440)^+$ signal in
$J/\psi$ photoproduction would be more challenging than
expected based on the LHCb result.

An evidence for $P_c^+$ was also found
in $\Lambda_b^0\to J/\psi p \pi^-$.
The $M_{J/\psi p}$ distribution [Fig.~3(b) of \cite{Pc_lhcb2}]
seems that
the $M_{J/\psi p}$ bin of $P_c(4440)^+$ is enhanced,
whereas this is not the case for the other $P_c^+$'s.
This observation may conflict with some $P_c^+$ models but
not ours because:
the $P_c(4440)^+$
[DT] mechanisms in $\Lambda_b^0\to J/\psi pK^-$
can [cannot] be shared by $\Lambda_b^0\to J/\psi p \pi^-$
and, thus, only $P_c(4440)^+$ appears in the latter.
For the limited statistics,
$P_c^+$ signals in $\Lambda_b^0\to J/\psi p \pi^-$ are still inconclusive.
Higher statistics data can seriously test the models.

\section{Summary and future perspective}

We analyzed the LHCb data for the $M_{J/\psi p}$ distribution of 
$\Lambda_b^0\to J/\psi p K^-$.
We found that the $P_c^+$ structures are well described by the double
triangle cusps and their interference with the common mechanisms. 
Only $P_c(4440)^+$ is interpreted as a resonance but its 
width and strength are significantly smaller than the LHCb result.
The analysis thus proposes an
understanding of the $P_c^+$ peaks completely different 
from the previous ones such as hadron molecules
and compact pentaquarks. 

An interesting next step is to study 
$\Lambda_b^0\to \Sigma_c^{(*)}\bar{D}^{(*)} K^-$.
An immediate prediction 
is TS peaks at $M_{\Sigma_c K^-}\sim 3.40$~GeV and
$M_{\Sigma_c^{*} K^-}\sim 3.21$~GeV due to
a single triangle diagram obtained by removing 
$\Sigma_c^{(*)}\bar{D}^{(*)}\to J/\psi p$ vertex 
from Fig.~\ref{fig:diag}(a).
To examine the $P_c^+$ structures,
a coupled-channel treatment of 
the $Y_c\bar{D}^{(*)}$ scattering 
is necessary.
It is also interesting to understand
other resonancelike structures near thresholds
with DT cusps that should now be in
the list of possible interpretations.

\begin{acknowledgments}
I thank Feng-Kun Guo and 
Christoph Hanhart for useful discussions and comments on the manuscript.
This work is in part supported by 
National Natural Science Foundation of China (NSFC) under contracts 
U2032103 and 11625523, 
and also by
National Key Research and Development Program of China under Contracts 2020YFA0406400.
\end{acknowledgments}

\begin{center}
 {\bf Supplemental material}
\end{center}

\setcounter{equation}{9}
\setcounter{figure}{3}

\begin{center}
 {1. $\Lambda_b^0\to J/\psi pK^-$ amplitudes}
\end{center}

\begin{figure}[t]
\begin{center}
\includegraphics[width=.5\textwidth]{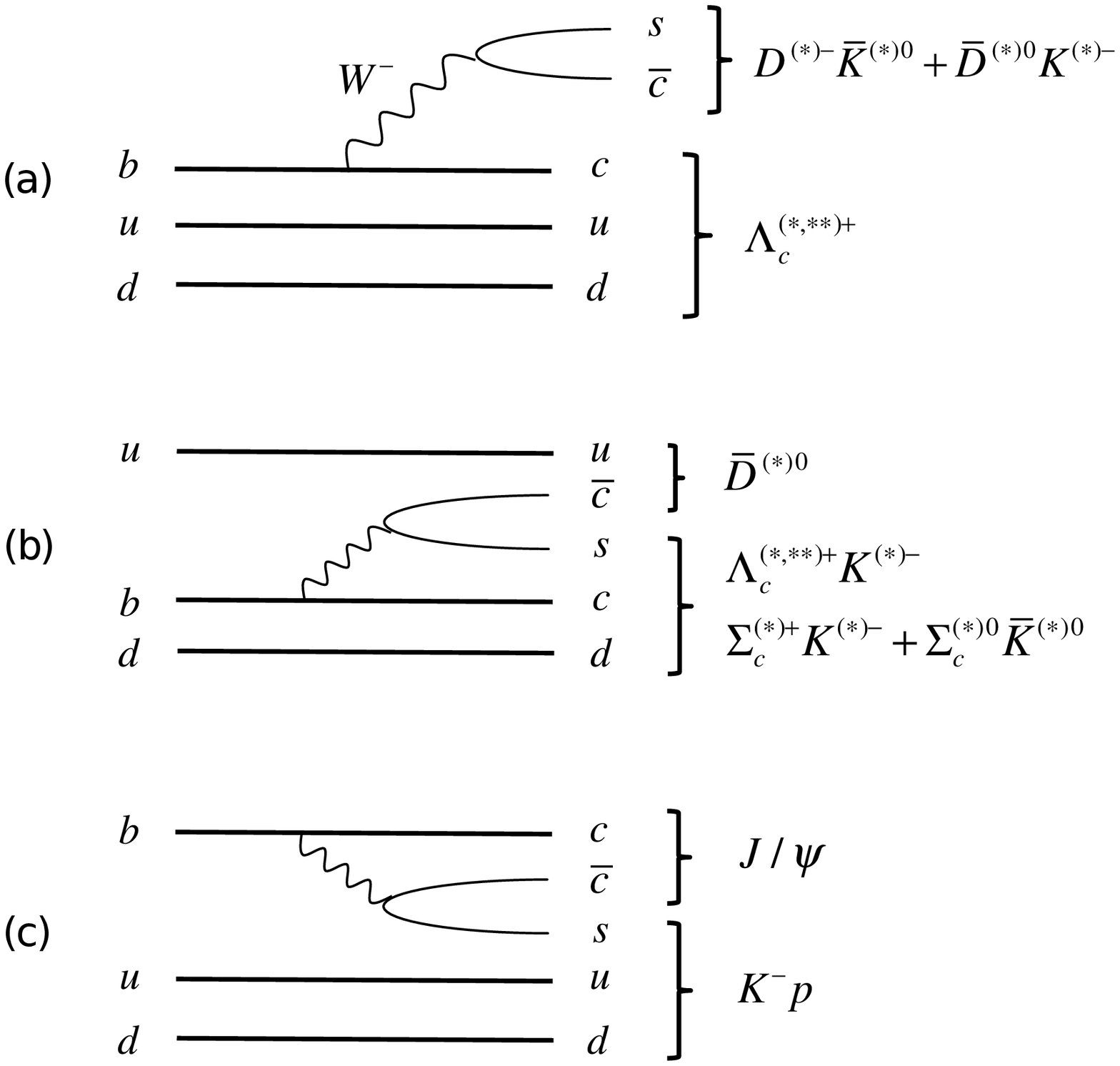}
\end{center}
 \caption{Quark diagrams for $\Lambda_b^0$ decay vertices.
(a) Color-favored diagram; (b,c) Color-suppressed diagrams.
 }
\label{fig:quark}
\end{figure}

We present formulas for diagrams of Fig.~1.
For the present study, we consider only parity-conserving mechanisms.
For double triangle (DT) diagrams of Fig.~1(a),
we consider those including 
$\Sigma_c\bar{D}(1/2^-)$,
$\Sigma^*_c\bar{D}(3/2^-)$, 
$\Sigma_c\bar{D}^*(1/2^-)$,
$\Sigma_c\bar{D}^*(3/2^-)$,
$\Sigma^*_c\bar{D}^*(1/2^-)$, and
$\Sigma^*_c\bar{D}^*(3/2^-)$
pairs going to $J/\psi p$
[$\Sigma_c\equiv\Sigma_c(2455)^{+,++}$,
$\Sigma_c^*\equiv\Sigma_c(2520)^{+,++}$].
Each of the DT diagrams 
includes four vertices.
The initial one
is either 
$\Lambda_b^0\to \Lambda_c^+\bar{D}\bar{K}^{*}$
or 
$\Lambda_b^0\to \Lambda_c^+\bar{D}^{*}\bar{K}^{*}$ vertices given by
\begin{eqnarray}
\label{eq:dt11}
&& c_{\Lambda_c\bar{D}\bar{K}^{*},\Lambda_b}
\bigg({1\over 2} t_{\bar{D}} {1\over 2} t_{\bar{K}^*}\bigg|00\bigg)\,
\bm{\sigma}\cdot\bm{\epsilon}_{\bar{K}^*}\,
 F_{\bar{K}^*\bar{D}\Lambda_c,\Lambda_b}^{00}, \\
\label{eq:dt12}
&& c_{\Lambda_c\bar{D}^*\bar{K}^{*},\Lambda_b}
\bigg({1\over 2} t_{\bar{D}^*} {1\over 2} t_{\bar{K}^*}\bigg|00\bigg)\,
i\bm{\sigma}\cdot
(\bm{\epsilon}_{\bar{K}^*}\times \bm{\epsilon}_{\bar{D}^*})\,
 F_{\bar{K}^*\bar{D}^*\Lambda_c,\Lambda_b}^{00},
\nonumber \\
\end{eqnarray}
respectively,
where the coupling constants 
$c_{\Lambda_c\bar{D}\bar{K}^{*},\Lambda_b}$ and
$c_{\Lambda_c\bar{D}^*\bar{K}^{*},\Lambda_b}$
are generally complex values.
Here and later in Eqs.~(\ref{eq:1L1i})-(\ref{eq:1L3i}),
we assume that color-favored $\Lambda_b^0$ decays 
[Fig.~\ref{fig:quark}(a)] dominate over 
color-suppressed ones [Fig.~\ref{fig:quark}(b)],
and thus the isospin of 
the $\bar{D}^{(*)}\bar{K}^{(*)}$ pair is 0.
We have used dipole form factors $F_{ijk,l}^{LL'}$ given as
\begin{eqnarray}
\label{eq:ff1}
 F_{ijk,l}^{LL'} =
 {1\over \sqrt{E_i E_j E_k E_l}}
\left(\frac{\Lambda^2}{\Lambda^2+q_{ij}^2}\right)^{\!\!2+{L\over 2}}\!\!\!
\left(\frac{\Lambda^{\prime 2}}{\Lambda^{\prime 2}+\tilde{p}_k^2}\right)^{\!\!2+{L'\over 2}}\!\!\!\!\!\!\!\!\!,\nonumber\\
\end{eqnarray}
where $q_{ij}$ ($\tilde{p}_{k}$) is the momentum of $i$ ($k$) in the
$ij$ (total) center-of-mass frame.
The second vertex, $\bar{K}^*\to\bar{K}\pi$ is common for all the
DT diagrams, and is given as
\begin{eqnarray}
 &&c_{\bar{K}\pi,\bar{K}^*}\, 
\bigg(1 t_\pi {1\over 2} t_{\bar{K}}\bigg| {1\over 2} t_{\bar{K}^*}\bigg)\,
\bm{\epsilon}_{\bar{K}^*} \cdot ( \bm{p}_{\bar{K}}-\bm{p}_{\pi})\,
 f_{\bar{K}\pi,\bar{K}^*}^{1}  ,
\end{eqnarray}
with the form factor $f_{ij,k}^{L}\equiv f_{ij}^{L}/\sqrt{E_k}$ and 
\begin{eqnarray}
 f_{ij}^{L} =
 {1\over \sqrt{E_i E_j}}
\left(\frac{\Lambda^2}{\Lambda^2+q_{ij}^2}\right)^{2+(L/2)}\ .
\label{eq:ff2}
\end{eqnarray}
The third vertex is either
$\Lambda_c\pi\to\Sigma_c$ or
$\Lambda_c\pi\to\Sigma^*_c$
which are given by
\begin{eqnarray}
&&c_{\Lambda_c\pi,\Sigma_c}\, 
\bm{\sigma} \cdot \bm{p}_{\pi} \,
 f_{\Lambda_c\pi,\Sigma_c}^{1}
\ , \\
&&c_{\Lambda_c\pi,\Sigma^*_c}\, 
\bm{S}^\dagger \cdot \bm{p}_{\pi} \,
 f_{\Lambda_c\pi,\Sigma^*_c}^{1}
\ , 
\end{eqnarray}
respectively.
We have introduced baryon spin operators
$\bm{S}^\dagger$ and $\bm{S}$ that 
change a baryon spin as 
${1\over 2}\to {3\over 2}$ and 
${3\over 2}\to {1\over 2}$, respectively;
they can be expressed with the Pauli matrices as
$\bm{S}\cdot\bm{a}\, \bm{S}^\dagger\cdot\bm{b}
={2\over 3} \bm{a}\cdot\bm{b}-{i\over 3} \bm{\sigma}\cdot 
(\bm{a}\times\bm{b})$.
The fourth vertices 
cause transitions
$\Sigma_c\bar{D}(1/2^-)$,
$\Sigma^*_c\bar{D}(3/2^-)$,
$\Sigma_c\bar{D}^*(1/2^-)$, 
$\Sigma_c\bar{D}^*(3/2^-)$, 
$\Sigma^*_c\bar{D}^*(1/2^-)$, 
$\Sigma^*_c\bar{D}^*(3/2^-) \to J/\psi p$ as given by
\begin{eqnarray}
&& c^{1/2^-}_{\psi p, \Sigma_c\bar{D}}
\bigg(1 t_{\Sigma_c} {1\over 2} t_{\bar{D}}\bigg| {1\over 2} t_p\bigg)
\bm{\sigma}\cdot \bm{\epsilon}_\psi \,
 f_{\psi p}^{0}\,
 f_{\Sigma_c\bar{D}}^{0}  , \\ 
&& c^{3/2^-}_{\psi p, \Sigma^*_c\bar{D}}
\bigg(1 t_{\Sigma^*_c} {1\over 2} t_{\bar{D}}\bigg| {1\over 2} t_p\bigg)
\bm{S}\cdot \bm{\epsilon}_\psi \,
 f_{\psi p}^{0}\,
 f_{\Sigma^*_c\bar{D}}^{0}  , \\ 
&& c^{1/2^-}_{\psi p, \Sigma_c\bar{D}^*}
\bigg(1 t_{\Sigma_c} {1\over 2} t_{\bar{D}^*}\bigg| {1\over 2} t_p\bigg)
\bm{\sigma}\!\cdot\! \bm{\epsilon}_\psi \,
\bm{\sigma}\!\cdot\! \bm{\epsilon}_{\bar{D}^*} \,
 f_{\psi p}^{0}\,
 f_{\Sigma_c\bar{D}^*}^{0}  , \\
&& c^{3/2^-}_{\psi p, \Sigma_c\bar{D}^*}
\bigg(1 t_{\Sigma_c} {1\over 2} t_{\bar{D}^*}\bigg| {1\over 2} t_p\bigg)\,
\bm{S}\!\cdot\! \bm{\epsilon}_\psi \,
\bm{S}^\dagger\!\cdot\! \bm{\epsilon}_{\bar{D}^*} 
 f_{\psi p}^{0}\,
 f_{\Sigma_c\bar{D}^*}^{0}  , \\
&& c^{1/2^-}_{\psi p, \Sigma^*_c\bar{D}^*}
\bigg(1 t_{\Sigma^*_c} {1\over 2} t_{\bar{D}^*}\bigg| {1\over 2} t_p\bigg)
\bm{\sigma}\!\cdot\! \bm{\epsilon}_\psi \,
\bm{S}\!\cdot\! \bm{\epsilon}_{\bar{D}^*} \,
 f_{\psi p}^{0}\,
 f_{\Sigma^*_c\bar{D}^*}^{0}  , \\
&& c^{3/2^-}_{\psi p, \Sigma^*_c\bar{D}^*}
\bigg(1 t_{\Sigma^*_c} {1\over 2} t_{\bar{D}^*}\bigg| {1\over 2} t_p\bigg)
\bm{S}\!\cdot\! \bm{\epsilon}_\psi \,
\bm{S}_{3\over 2}\!\!\cdot\! \bm{\epsilon}_{\bar{D}^*} 
 f_{\psi p}^{0}\,
 f_{\Sigma^*_c\bar{D}^*}^{0}  , 
\end{eqnarray}
respectively;
$\bm{S}_{3\over 2}$ is a spin operator defined with Clebsch-Gordan
coefficient as 
$\bra{m'}S_{3\over 2}^k\ket{m}\equiv ({3\over 2} m 1 k\, |\, {3\over 2} m')$
where $\ket{m^{(\prime)}}$ is a spin state of a spin-3/2 particle.
The DT amplitudes
including $\Sigma^{(*)}_c\bar{D}^{(*)}(J^P)$,
denoted by $A^{\rm DT}_{\Sigma^{(*)}_c\bar{D}^{(*)}(J^P)}$,
are constructed with the above ingredients as

\begin{widetext}

\begin{eqnarray}
 A^{\rm DT}_{\Sigma_c\bar{D}(1/2^-)} &=&
c^{1/2^-}_{\psi p, \Sigma_c\bar{D}}\,
c_{\Lambda_c\pi,\Sigma_c}\, 
c_{\bar{K}\pi,\bar{K}^*}\, 
c_{\Lambda_c\bar{D}\bar{K}^{*},\Lambda_b}
\bigg(1 t_{\Sigma_c} {1\over 2} t_{\bar{D}}\bigg| {1\over 2} t_p\bigg)
\bigg(1 t_\pi {1\over 2} t_{\bar{K}}\bigg| {1\over 2} t_{\bar{K}^*}\bigg)
\bigg({1\over 2} t_{\bar{D}} {1\over 2} t_{\bar{K}^*}\bigg|00\bigg)
\!\int\! d^3p_{\pi}\!\int\! d^3p_{\bar{D}}
\nonumber\\
&&\times
{\bm{\sigma}\cdot \bm{\epsilon}_\psi\,
\bm{\sigma}\cdot \bm{p}_\pi \,
\over E-E_{\bar{K}}-E_{\Sigma_c}-E_{\bar{D}}+{i\over 2}\Gamma_{\Sigma_c}}
{\bm{\sigma}\cdot (\bm{p}_{\bar{K}}-\bm{p}_\pi)
\over E-E_{\bar{K}}-E_\pi-E_{\Lambda_c}-E_{\bar{D}}+i\epsilon}
{f_{\psi p}^{0}\,
 f_{\Sigma_c\bar{D}}^{0}
 f_{\Lambda_c\pi,\Sigma_c}^{1}
 f_{\bar{K}\pi,\bar{K}^*}^{1}
 F_{\bar{K}^*\bar{D}\Lambda_c,\Lambda_b}^{00}
\over E-E_{\bar{K}^*}-E_{\Lambda_c}-E_{\bar{D}}+{i\over 2}\Gamma_{K^*}},
\label{eq:DT2}
\\
 A^{\rm DT}_{\Sigma^*_c\bar{D}(3/2^-)} &=&
c^{3/2^-}_{\psi p, \Sigma^*_c\bar{D}}\,
c_{\Lambda_c\pi,\Sigma^*_c}\, 
c_{\bar{K}\pi,\bar{K}^*}\, 
c_{\Lambda_c\bar{D}\bar{K}^{*},\Lambda_b}
\bigg(1 t_{\Sigma^*_c} {1\over 2} t_{\bar{D}}\bigg| {1\over 2} t_p\bigg)
\bigg(1 t_\pi {1\over 2} t_{\bar{K}}\bigg| {1\over 2} t_{\bar{K}^*}\bigg)
\bigg({1\over 2} t_{\bar{D}} {1\over 2} t_{\bar{K}^*}\bigg|00\bigg)
\!\int\! d^3p_{\pi}\!\int\! d^3p_{\bar{D}}
\nonumber\\
&&\times
{\bm{S}\cdot \bm{\epsilon}_\psi\,
\bm{S}^\dagger\cdot \bm{p}_\pi \,
\over E-E_{\bar{K}}-E_{\Sigma^*_c}-E_{\bar{D}}+{i\over 2}\Gamma_{\Sigma^*_c}}
{
\bm{\sigma}
\cdot (\bm{p}_{\bar{K}}-\bm{p}_\pi)
\over E-E_{\bar{K}}-E_\pi-E_{\Lambda_c}-E_{\bar{D}}+i\epsilon}
{f_{\psi p}^{0}\,
 f_{\Sigma^*_c\bar{D}}^{0}
 f_{\Lambda_c\pi,\Sigma^*_c}^{1}
 f_{\bar{K}\pi,\bar{K}^*}^{1}
 F_{\bar{K}^*\bar{D}\Lambda_c,\Lambda_b}^{00}
\over E-E_{\bar{K}^*}-E_{\Lambda_c}-E_{\bar{D}}+{i\over 2}\Gamma_{K^*}},
\label{eq:DT12}
\\
 A^{\rm DT}_{\Sigma_c\bar{D}^*(1/2^-)} &=&
c^{1/2^-}_{\psi p, \Sigma_c\bar{D}^*}\,
c_{\Lambda_c\pi,\Sigma_c}\, 
c_{\bar{K}\pi,\bar{K}^*}\, 
c_{\Lambda_c\bar{D}^*\bar{K}^{*},\Lambda_b}
\bigg(1 t_{\Sigma_c} {1\over 2} t_{\bar{D}^*}\bigg| {1\over 2} t_p\bigg)
\bigg(1 t_\pi {1\over 2} t_{\bar{K}}\bigg| {1\over 2} t_{\bar{K}^*}\bigg)
\bigg({1\over 2} t_{\bar{D}^*} {1\over 2} t_{\bar{K}^*}\bigg|00\bigg)
\!\int\! d^3p_{\pi}\!\int\! d^3p_{\bar{D}^*}
\nonumber\\
&&\times
{\sum_{\epsilon_{\bar{D}^*}}
\bm{\sigma}\cdot \bm{\epsilon}_\psi\,
\bm{\sigma}\cdot \bm{\epsilon}_{\bar{D}^*} \,
\bm{\sigma}\cdot \bm{p}_\pi \,
\over E-E_{\bar{K}}-E_{\Sigma_c}-E_{\bar{D}^*}+{i\over 2}\Gamma_{\Sigma_c}}
{
i\bm{\sigma}\cdot \left[(\bm{p}_{\bar{K}}-\bm{p}_\pi)\times \bm{\epsilon}_{\bar{D}^*}\right]
\over E-E_{\bar{K}}-E_\pi-E_{\Lambda_c}-E_{\bar{D}^*}+i\epsilon}
{f_{\psi p}^{0}\,
 f_{\Sigma_c\bar{D}^*}^{0}
 f_{\Lambda_c\pi,\Sigma_c}^{1}
 f_{\bar{K}\pi,\bar{K}^*}^{1}
 F_{\bar{K}^*\bar{D}^*\Lambda_c,\Lambda_b}^{00}
\over E-E_{\bar{K}^*}-E_{\Lambda_c}-E_{\bar{D}^*}+{i\over 2}\Gamma_{K^*}}, \nonumber\\ 
\label{eq:DT13}
\\
 A^{\rm DT}_{\Sigma_c\bar{D}^*(3/2^-)} &=&
c^{3/2^-}_{\psi p, \Sigma_c\bar{D}^*}\,
c_{\Lambda_c\pi,\Sigma_c}\, 
c_{\bar{K}\pi,\bar{K}^*}\, 
c_{\Lambda_c\bar{D}^*\bar{K}^{*},\Lambda_b}
\bigg(1 t_{\Sigma_c} {1\over 2} t_{\bar{D}^*}\bigg| {1\over 2} t_p\bigg)
\bigg(1 t_\pi {1\over 2} t_{\bar{K}}\bigg| {1\over 2} t_{\bar{K}^*}\bigg)
\bigg({1\over 2} t_{\bar{D}^*} {1\over 2} t_{\bar{K}^*}\bigg|00\bigg)
\!\int\! d^3p_{\pi}\!\int\! d^3p_{\bar{D}^*}
\nonumber\\
&&\times
{\sum_{\epsilon_{\bar{D}^*}}
\bm{S}\cdot \bm{\epsilon}_\psi\,
\bm{S}^\dagger\cdot \bm{\epsilon}_{\bar{D}^*} \,
\bm{\sigma}\cdot \bm{p}_\pi \,
\over E-E_{\bar{K}}-E_{\Sigma_c}-E_{\bar{D}^*}+{i\over 2}\Gamma_{\Sigma_c}}
{
i\bm{\sigma}\cdot \left[(\bm{p}_{\bar{K}}-\bm{p}_\pi)\times \bm{\epsilon}_{\bar{D}^*}\right]
\over E-E_{\bar{K}}-E_\pi-E_{\Lambda_c}-E_{\bar{D}^*}+i\epsilon}
{f_{\psi p}^{0}\,
 f_{\Sigma_c\bar{D}^*}^{0}
 f_{\Lambda_c\pi,\Sigma_c}^{1}
 f_{\bar{K}\pi,\bar{K}^*}^{1}
 F_{\bar{K}^*\bar{D}^*\Lambda_c,\Lambda_b}^{00}
\over E-E_{\bar{K}^*}-E_{\Lambda_c}-E_{\bar{D}^*}+{i\over 2}\Gamma_{K^*}}, \nonumber\\
\label{eq:DT14}
\\
 A^{\rm DT}_{\Sigma^*_c\bar{D}^*(1/2^-)} &=&
c^{1/2^-}_{\psi p, \Sigma^*_c\bar{D}^*}\,
c_{\Lambda_c\pi,\Sigma^*_c}\, 
c_{\bar{K}\pi,\bar{K}^*}\, 
c_{\Lambda_c\bar{D}^*\bar{K}^{*},\Lambda_b}
\bigg(1 t_{\Sigma^*_c} {1\over 2} t_{\bar{D}^*}\bigg| {1\over 2} t_p\bigg)
\bigg(1 t_\pi {1\over 2} t_{\bar{K}}\bigg| {1\over 2} t_{\bar{K}^*}\bigg)
\bigg({1\over 2} t_{\bar{D}^*} {1\over 2} t_{\bar{K}^*}\bigg|00\bigg)
\!\int\! d^3p_{\pi}\!\int\! d^3p_{\bar{D}^*}
\nonumber\\
&&\times
{\sum_{\epsilon_{\bar{D}^*}}
\bm{\sigma}\cdot \bm{\epsilon}_\psi\,
\bm{S}\cdot \bm{\epsilon}_{\bar{D}^*} \,
\bm{S}^\dagger\cdot \bm{p}_\pi \,
\over E-E_{\bar{K}}-E_{\Sigma^*_c}-E_{\bar{D}^*}+{i\over 2}\Gamma_{\Sigma^*_c}}
{
i\bm{\sigma}\cdot \left[(\bm{p}_{\bar{K}}-\bm{p}_\pi)\times \bm{\epsilon}_{\bar{D}^*}\right]
\over E-E_{\bar{K}}-E_\pi-E_{\Lambda_c}-E_{\bar{D}^*}+i\epsilon}
{f_{\psi p}^{0}\,
 f_{\Sigma^*_c\bar{D}^*}^{0}
 f_{\Lambda_c\pi,\Sigma^*_c}^{1}
 f_{\bar{K}\pi,\bar{K}^*}^{1}
 F_{\bar{K}^*\bar{D}^*\Lambda_c,\Lambda_b}^{00}
\over E-E_{\bar{K}^*}-E_{\Lambda_c}-E_{\bar{D}^*}+{i\over 2}\Gamma_{K^*}}, \nonumber\\
\label{eq:DT15}
%
\end{eqnarray}
\begin{eqnarray}
 A^{\rm DT}_{\Sigma^*_c\bar{D}^*(3/2^-)} &=&
c^{3/2^-}_{\psi p, \Sigma^*_c\bar{D}^*}\,
c_{\Lambda_c\pi,\Sigma^*_c}\, 
c_{\bar{K}\pi,\bar{K}^*}\, 
c_{\Lambda_c\bar{D}^*\bar{K}^{*},\Lambda_b}
\bigg(1 t_{\Sigma^*_c} {1\over 2} t_{\bar{D}^*}\bigg| {1\over 2} t_p\bigg)
\bigg(1 t_\pi {1\over 2} t_{\bar{K}}\bigg| {1\over 2} t_{\bar{K}^*}\bigg)
\bigg({1\over 2} t_{\bar{D}^*} {1\over 2} t_{\bar{K}^*}\bigg|00\bigg)
\!\int\! d^3p_{\pi}\!\int\! d^3p_{\bar{D}^*}
\nonumber\\
&&\times
{\sum_{\epsilon_{\bar{D}^*}}
\bm{S}\cdot \bm{\epsilon}_\psi\,
\bm{S}_{3/2}\cdot \bm{\epsilon}_{\bar{D}^*} \,
\bm{S}^\dagger\cdot \bm{p}_\pi \,
\over E-E_{\bar{K}}-E_{\Sigma^*_c}-E_{\bar{D}^*}+{i\over 2}\Gamma_{\Sigma^*_c}}
{
i\bm{\sigma}\cdot \left[(\bm{p}_{\bar{K}}-\bm{p}_\pi)\times \bm{\epsilon}_{\bar{D}^*}\right]
\over E-E_{\bar{K}}-E_\pi-E_{\Lambda_c}-E_{\bar{D}^*}+i\epsilon}
{f_{\psi p}^{0}\,
 f_{\Sigma^*_c\bar{D}^*}^{0}
 f_{\Lambda_c\pi,\Sigma^*_c}^{1}
 f_{\bar{K}\pi,\bar{K}^*}^{1}
 F_{\bar{K}^*\bar{D}^*\Lambda_c,\Lambda_b}^{00}
\over E-E_{\bar{K}^*}-E_{\Lambda_c}-E_{\bar{D}^*}+{i\over 2}\Gamma_{K^*}},\nonumber\\
\label{eq:DT3}
\end{eqnarray}
\end{widetext}
where, in each amplitude, the summation over the two charged states
with the charge dependent particle masses is implicit;
the tiny $\Gamma_{D^{*}}$ has been neglected.
It is understood that the amplitudes are implicitly sandwiched by 
initial $\Lambda_b^0$ and final $p$ spin state.
Next we present formulas for 
the one-loop diagrams
of Fig.~1(b)
including the $\Lambda_c^+\bar{D}^{*0}(1/2^-)$,
$\Lambda_c(2593)^+\bar{D}^{0}(1/2^+)$, and
$\Lambda_c(2625)^+\bar{D}^{0}(3/2^+)$ loops.
The initial decay vertices are:
\begin{eqnarray}
\label{eq:1L1i}
&& c_{\Lambda_c\bar{D}^*\bar{K},\Lambda_b}
\bigg({1\over 2} t_{\bar{D}^*} {1\over 2} t_{\bar{K}}\bigg|00\bigg)\,
\bm{\sigma}\cdot\bm{\epsilon}_{\bar{D}^*}
 F_{\bar{K}\bar{D}^*\Lambda_c,\Lambda_b}^{00}\ , \\
&& c_{\Lambda^*_c\bar{D}\bar{K},\Lambda_b}
\bigg({1\over 2} t_{\bar{D}} {1\over 2} t_{\bar{K}}\bigg|00\bigg)\,
\bm{\sigma}
\cdot (\bm{p}_{\bar{D}} - \bm{p}_{\bar{K}})
F_{\bar{K}\bar{D}\Lambda^*_c,\Lambda_b}^{10}\, , 
\end{eqnarray}
\begin{eqnarray}
\label{eq:1L3i}
&& c_{\Lambda^{**}_c\bar{D}\bar{K},\Lambda_b}
\bigg({1\over 2} t_{\bar{D}} {1\over 2} t_{\bar{K}}\bigg|00\bigg)\,
\bm{S}^\dagger\cdot  (\bm{p}_{\bar{D}} - \bm{p}_{\bar{K}})
 F_{\bar{K}\bar{D}\Lambda^{**}_c,\Lambda_b}^{10}\ , 
\nonumber \\
\end{eqnarray}
for $\Lambda_b^0\to \Lambda_c^+\bar{D}^{*0}K^-, \Lambda^{*+}_c\bar{D}^{0}K^-$,
and $\Lambda^{**+}_c\bar{D}^{0}K^-$, respectively
[$\Lambda^*_c\equiv\Lambda_c(2595)^+$, $\Lambda^{**}_c\equiv\Lambda_c(2625)^+$].
The coupling constants 
$c_{\Lambda^{(*,**)}_c\bar{D}^{(*)}\bar{K},\Lambda_b}$
are generally complex values.
The subsequent interactions for 
$\Lambda_c^{+}\bar{D}^{*0}(1/2^-)$,
$\Lambda_c^{*+}\bar{D}^{0}(1/2^+)$,
$\Lambda_c^{**+}\bar{D}^{0}(3/2^+)\to J/\psi p$ are
\begin{eqnarray}
&& c^{1/2^-}_{\psi p, \Lambda_c\bar{D}^*}
\bm{\sigma}\cdot \bm{\epsilon}_\psi \,
\bm{\sigma}\cdot\bm{\epsilon}_{\bar{D}^*}
 f_{\psi p}^{0}
 f_{\Lambda_c\bar{D}^*}^{0} \ , \\
&& c^{1/2^+}_{\psi p, \Lambda^*_c\bar{D}}
\bm{\sigma}\cdot \bm{\epsilon}_\psi 
\bm{\sigma}\cdot \bm{p}_\psi
 f_{\psi p}^{1}
 f_{\Lambda^*_c\bar{D}}^{0} \ , \\
&& c^{3/2^+}_{\psi p, \Lambda^{**}_c\bar{D}}
\bm{\sigma}\cdot \bm{\epsilon}_\psi 
\bm{S}\cdot \bm{p}_\psi
 f_{\psi p}^{1}
 f_{\Lambda^{**}_c\bar{D}}^{0} \ ,
\end{eqnarray}
respectively.
The one-loop amplitudes 
including $\Lambda_c^{(*,**)+}\bar{D}^{(*)0}$,
denoted by $A^{\rm 1L}_{\Lambda_c^{(*,**)}\bar{D}^{(*)}}$,
are given with the above ingredients as
\begin{widetext}
\begin{eqnarray}
\label{eq:1L2}
 A^{\rm 1L}_{\Lambda_c\bar{D}^*} &=&
3 \, 
 c^{1/2^{-}}_{\psi p, \Lambda_c\bar{D}^*}\,
 c_{\Lambda_c\bar{D}^*\bar{K},\Lambda_b}\,
\bigg({1\over 2} t_{\bar{D}^*} {1\over 2} t_{\bar{K}}\bigg|00\bigg)\,
\bm{\sigma}\cdot \bm{\epsilon}_\psi 
\int d^3p_{\bar{D}^*} 
 {
 f_{\psi p}^{0}
 f_{\Lambda_c\bar{D}^*}^{0}
 F_{\bar{K}\bar{D}^*\Lambda_c,\Lambda_b}^{00}
\over E-E_{\bar{K}}-E_{\Lambda_c}-E_{\bar{D}^*}+{i\over 2}\Gamma_{D^*}}, \\
\label{eq:1L12}
 A^{\rm 1L}_{\Lambda^*_c\bar{D}} &=&
 c^{1/2^{+}}_{\psi p, \Lambda^*_c\bar{D}}\,
 c_{\Lambda^*_c\bar{D}\bar{K},\Lambda_b}\,
\bigg({1\over 2} t_{\bar{D}} {1\over 2} t_{\bar{K}}\bigg|00\bigg)\,
\bm{\sigma}\cdot \bm{\epsilon}_\psi \,
\bm{\sigma}\cdot \bm{p}_\psi
\int d^3p_{\bar{D}} \,
\bm{\sigma}
\cdot (\bm{p}_{\bar{D}}- \bm{p}_{\bar{K}})
 { 
 f_{\psi p}^{1}
 f_{\Lambda^*_c\bar{D}}^{0}
 F_{\bar{K}\bar{D}\Lambda^*_c,\Lambda_b}^{10}
\over E-E_{\bar{K}}-E_{\Lambda^*_c}-E_{\bar{D}}+{i\over 2}\Gamma_{\Lambda^*_c}}
\ , \\
 A^{\rm 1L}_{\Lambda^{**}_c\bar{D}} &=&
 c^{3/2^{+}}_{\psi p, \Lambda^{**}_c\bar{D}}\,
 c_{\Lambda^{**}_c\bar{D}\bar{K},\Lambda_b}
\bigg({1\over 2} t_{\bar{D}} {1\over 2} t_{\bar{K}}\bigg|00\bigg)
\bm{\sigma}\!\cdot\! \bm{\epsilon}_\psi 
\bm{S}\!\cdot\! \bm{p}_\psi
\!\int\! d^3p_{\bar{D}} \
\bm{S}^\dagger\!\!\cdot\!  (\bm{p}_{\bar{D}} - \bm{p}_{\bar{K}})
 { f_{\psi p}^{1}
 f_{\Lambda^{**}_c\bar{D}}^{0} 
 F_{\bar{K}\bar{D}\Lambda^{**}_c,\Lambda_b}^{10}
\over E-E_{\bar{K}}-E_{\Lambda^{**}_c}-E_{\bar{D}}+{i\over 2}\Gamma_{\Lambda^{**}_c}}.
\label{eq:1L3}
\end{eqnarray}
\end{widetext}

The resonant $P_c(4440)^+$ amplitude
of Fig.~1(c) is given as
\begin{eqnarray}
\label{eq:pc1hf}
 A^{1/2^{-}}_{P_c(4440)} &=&
c^{1/2^{-}}_{P_c(4440)}\,
{\bm{\sigma}\cdot \bm{\epsilon}_\psi\,
f_{\psi p,P_c}^{0}
f_{P_c\bar{K},\Lambda_b}^{0}
\over E-E_{\bar{K}}-E_{P_c}+{i\over 2}\Gamma_{P_c}}
\ , 
\end{eqnarray}
for $P_c(4440)^+$ of $J^P=1/2^-$,
\begin{eqnarray}
\label{eq:pc3hf}
 A^{3/2^{-}}_{P_c(4440)} &=&
c^{3/2^{-}}_{P_c(4440)}\,
\bm{S}\cdot \bm{\epsilon}_\psi\,
\bm{S}^\dagger\cdot \bm{p}_{\bar{K}}\,
\bm{\sigma}\cdot \bm{p}_{\bar{K}}\,
 \nonumber \\
&&\times 
{ 
f_{\psi p,P_c}^{0}
f_{P_c\bar{K},\Lambda_b}^{2}
\over E-E_{\bar{K}}-E_{P_c}+{i\over 2}\Gamma_{P_c}},
\end{eqnarray}
for $P_c(4440)^+$ of $J^P=3/2^-$, 
\begin{eqnarray}
\label{eq:pc1hfp}
 A^{1/2^{+}}_{P_c(4440)} &=&
c^{1/2^{+}}_{P_c(4440)}\,
\bm{\sigma}\cdot \bm{\epsilon}_\psi\,
\bm{\sigma}\cdot \bm{p}_\psi\,
\bm{\sigma}\cdot \bm{p}_{\bar{K}}\,
 \nonumber \\
&&\times 
{
f_{\psi p,P_c}^{1}
f_{P_c\bar{K},\Lambda_b}^{1}
\over E-E_{\bar{K}}-E_{P_c}+{i\over 2}\Gamma_{P_c}}
\ , 
\end{eqnarray}
for $P_c(4440)^+$ of $J^P=1/2^+$, and 
\begin{eqnarray}
\label{eq:pc3hfp}
 A^{3/2^{+}}_{P_c(4440)} &=&
c^{3/2^{+}}_{P_c(4440)}\,
\bm{\sigma}\cdot \bm{\epsilon}_\psi\,
\bm{S}\cdot \bm{p}_\psi\,
\bm{S^\dagger}\cdot \bm{p}_{\bar{K}}\,
 \nonumber \\
&&\times 
{
f_{\psi p,P_c}^{1}
f_{P_c\bar{K},\Lambda_b}^{1}
\over E-E_{\bar{K}}-E_{P_c}+{i\over 2}\Gamma_{P_c}}
\ , 
\end{eqnarray}
for $P_c(4440)^+$ of $J^P=3/2^+$.

The direct decays [Fig.~\ref{fig:diag}(d)] 
may mainly 
originate from a color-suppressed quark diagram of
Fig.~\ref{fig:quark}(c).
The corresponding amplitudes can be projected onto the
$pJ/\psi(J^P)$ partial waves.
Thus we employ a form 
as follows:
\begin{eqnarray}
\label{eq:dir_s}
 A_{\rm dir} &=&
  c_{\rm dir}^{1/2^-}
\bm{\sigma}\cdot \bm{\epsilon}_\psi  \,
 F_{\psi p\bar{K},\Lambda_b}^{00} \nonumber
\\
&&
+ c_{\rm dir}^{3/2^-}\, 
\bm{S}\cdot \bm{\epsilon}_\psi\,
\bm{S}^\dagger\cdot \bm{p}_{\bar{K}}\,
\bm{\sigma}\cdot \bm{p}_{\bar{K}}\,
 F_{\psi p\bar{K},\Lambda_b}^{02} \nonumber\\
&&
+ c_{\rm dir}^{1/2^+}
\bm{\sigma}\cdot \bm{\epsilon}_\psi\,  
\bm{\sigma}\cdot \bm{p}_\psi  \,
\bm{\sigma}\cdot \bm{p}_{\bar{K}}\,
 F_{\psi p\bar{K},\Lambda_b}^{11} \nonumber\\
&&
+ c_{\rm dir}^{3/2^+}
\bm{\sigma}\cdot \bm{\epsilon}_\psi\,  
\bm{S}\cdot \bm{p}_\psi  \,
\bm{S}^\dagger\cdot \bm{p}_{\bar{K}}\,
 F_{\psi p\bar{K},\Lambda_b}^{11} ,
\end{eqnarray}
where $c_{\rm dir}^{J^P}$ is a (real) coupling constant
for the $pJ/\psi(J^P)$ partial wave amplitude;
the lowest orbital angular momentum between 
$pJ/\psi(J^P)$ and $K^-$ is considered.
We basically use the common cutoff value for all the vertices discussed
above.
One exception applies to Eq.~(\ref{eq:dir_s})
where we adjust $\Lambda^\prime$ of Eq.~(\ref{eq:ff1})
so that the $M_{J/\psi p}$ distribution from the direct decay amplitude
is similar to the phase-space shape.

In numerical calculations, 
for convenience, all the above amplitudes 
are evaluated in the $J/\psi p$ center-of-mass frame.
With the relevant kinematical factors multiplied,
the invariant amplitudes are obtained and are plugged into the Dalitz plot
distribution formula. See Appendix~B of Ref.~\cite{3pi} for details.

\begin{table}[b]
\renewcommand{\arraystretch}{1.8}
\tabcolsep=3.mm
\caption{\label{tab:para}
Parameter values for the full model ($\Lambda=1$~GeV)
obtained from fitting the LHCb data~\cite{lhcb_pc}.
Each of the parameters is included in the amplitude specified 
by the equation label in the third column. 
The parameters above the horizontal line can be arbitrarily scaled by 
a common overall factor. 
The parameter value in the last row is shared as
$h_{\Lambda_c\bar{D}^*(1/2^-)}=h_{\Sigma_c^*\bar{D}^*(1/2^-)}=h_{\Sigma_c^*\bar{D}^*(3/2^-)}$.
}
\begin{tabular}[b]{lrr}
$c^{1/2^-}_{\psi p, \Sigma_c\bar{D}}\, c_{\Lambda_c\bar{D}\bar{K}^{*},\Lambda_b}$      &$ -2.21 +4.79 \,i$ &Eq.~(\ref{eq:DT2})   \\
$c^{3/2^-}_{\psi p, \Sigma^*_c\bar{D}}\, c_{\Lambda_c\bar{D}\bar{K}^{*},\Lambda_b}$    &$  5.59 -7.49 \,i$ &Eq.~(\ref{eq:DT12})  \\
$c^{1/2^-}_{\psi p, \Sigma_c\bar{D}^*}\, c_{\Lambda_c\bar{D}^*\bar{K}^{*},\Lambda_b}$  &$ -3.43 -1.12 \,i$ &Eq.~(\ref{eq:DT13})  \\
$c^{3/2^-}_{\psi p, \Sigma_c\bar{D}^*}\, c_{\Lambda_c\bar{D}^*\bar{K}^{*},\Lambda_b}$  &$  0.20+10.46 \,i$ &Eq.~(\ref{eq:DT14})  \\
$c^{1/2^-}_{\psi p, \Sigma^*_c\bar{D}^*}\, c_{\Lambda_c\bar{D}^*\bar{K}^{*},\Lambda_b}$&$  5.00 -1.40 \,i$ &Eq.~(\ref{eq:DT15})  \\
$c^{3/2^-}_{\psi p, \Sigma^*_c\bar{D}^*}\, c_{\Lambda_c\bar{D}^*\bar{K}^{*},\Lambda_b}$&$ 12.84 +9.38 \,i$ &Eq.~(\ref{eq:DT3})   \\
$c^{1/2^{-}}_{\psi p, \Lambda_c\bar{D}^*}\,  c_{\Lambda_c\bar{D}^*\bar{K},\Lambda_b}$                  &$ -0.38 -0.63 \,i$&Eq.~(\ref{eq:1L2})   \\
$c^{1/2^{+}}_{\psi p, \Lambda^*_c\bar{D}}\,  c_{\Lambda^*_c\bar{D}\bar{K},\Lambda_b}$ (GeV$^{-2}$)     &$ -0.76 -0.24 \,i$&Eq.~(\ref{eq:1L12})  \\
$c^{3/2^{+}}_{\psi p, \Lambda^{**}_c\bar{D}}\,c_{\Lambda^{**}_c\bar{D}\bar{K},\Lambda_b}$ (GeV$^{-2}$) &$ -0.03 -1.47 \,i$&Eq.~(\ref{eq:1L3})   \\
$c^{3/2^{-}}_{P_c(4440)}\ \ (\times 10^{2})$ & $  4.31 -3.30 \,i$& Eq.~(\ref{eq:pc3hf})\\
$c_{\rm dir}^{1/2^-}$              &$   2.01$ & Eq.~(\ref{eq:dir_s})\\
$c_{\rm dir}^{3/2^-}$ (GeV$^{-2}$) &$-119.86$ & Eq.~(\ref{eq:dir_s}) \\
$c_{\rm dir}^{1/2^+}$ (GeV$^{-2}$) &$   5.69$ & Eq.~(\ref{eq:dir_s}) \\
$c_{\rm dir}^{3/2^+}$ (GeV$^{-2}$) &$   8.23$ & Eq.~(\ref{eq:dir_s}) \\\hline
$m_{P_c}$ (MeV)& 4443.1 & Eq.~(\ref{eq:pc3hf})\\
$\Gamma_{P_c}$ (MeV) & 2.7 & Eq.~(\ref{eq:pc3hf})\\
$h_{\Lambda_c\bar{D}^*(1/2^-)}$ & 2.50 &Eq.~(\ref{eq:cont-ptl})
\end{tabular}
\end{table}

\begin{table}[t]
\renewcommand{\arraystretch}{1.5}
\tabcolsep=4.5mm
\caption{\label{tab:para2}
Parameter values for the full model ($\Lambda=1$~GeV)
{\it not} fitted to the LHCb data~\cite{lhcb_pc}.
The parameter value in the last row is shared as
$h_{\Sigma_c\bar{D}(1/2^-)}=h_{\Sigma_c\bar{D}^*(1/2^-)}=h_{\Sigma_c\bar{D}^*(3/2^-)}=h_{\Sigma_c^*\bar{D}(3/2^-)}=
h_{\Lambda^*_c\bar{D}(1/2^+)}=h_{\Lambda^{**}_c\bar{D}(3/2^+)}$.
}
\begin{tabular}[b]{lrr}
$c_{\bar{K}\pi,\bar{K}^*}$   &$0.15$ &Eqs.~(\ref{eq:DT2})-(\ref{eq:DT3})  \\
$c_{\Lambda_c\pi,\Sigma_c}$  &$0.46$ &Eqs.~(\ref{eq:DT2}), (\ref{eq:DT13}), (\ref{eq:DT14})  \\
$c_{\Lambda_c\pi,\Sigma^*_c}$&$0.87$ &Eqs.~(\ref{eq:DT12}), (\ref{eq:DT15}), (\ref{eq:DT3})  \\
$\Lambda$ (MeV) &1000 &  \\
$\Lambda_{\rm dir}^{^\prime 1/2^-}$ (MeV) &800 &Eq.~(\ref{eq:dir_s})  \\
$\Lambda_{\rm dir}^{^\prime 3/2^-}$ (MeV) &400 &Eq.~(\ref{eq:dir_s})  \\
$\Lambda_{\rm dir}^{^\prime 1/2^+}$ (MeV) &800 &Eq.~(\ref{eq:dir_s})  \\
$\Lambda_{\rm dir}^{^\prime 3/2^+}$ (MeV) &800 &Eq.~(\ref{eq:dir_s})  \\
$h_{\Sigma_c\bar{D}(1/2^-)}$ & $-2$ &Eq.~(\ref{eq:cont-ptl})
\end{tabular}
\end{table}

We remark on the choice of the form factors.
We choose the dipole form without a particular reason.
Other choices such
as monopole and exponential forms could also be used, and they should
not significantly change the conclusion. This is because 
the kinematical effects that cause resonancelike structures are not
sensitive to the dynamical details such as a particular form of the
form factors. 
We varied the cutoff value over a rather wide range (0.8--2 GeV), and confirmed
the stability of the result. This large cutoff variation
essentially checked the stability of the result against changing the functional form
of the form factors.

The parameter values obtained from fitting and not from fitting the LHCb data~\cite{lhcb_pc}
are presented in Table~\ref{tab:para} and \ref{tab:para2}, respectively.
All  coupling parameters for the DT amplitudes have similar magnitudes. 
Although $c_{\rm dir}^{3/2^-}=-119.86$~GeV$^{-2}$ seems noticeably
larger than the others, this is partly due to the use of a small cutoff value 
$\Lambda_{\rm dir}^{^\prime 3/2^-}=400$~MeV.
Here, we remark on possible constraints
on the parameters
due to the heavy quark spin symmetry (HQSS).
The HQSS puts a relation
among $c^{J^P}_{\psi p, \Sigma^{(*)}_c\bar{D}^{(*)}}$
as discussed in Ref.~\cite{pc_du}.
However, this HQSS relation
might not strictly constrain the fits in this work.
This is because
$c^{J^P}_{\psi p, \Sigma^{(*)}_c\bar{D}^{(*)}}$ is always accompanied by
$c_{\Lambda_c\bar{D}^{(*)}\bar{K}^*,\Lambda_b}$,
and we can determine only their product
as seen in the table.
The HQSS constraint would be even less restrictive
considering that 
we could have also included:
(i) parity-violating initial vertices
with more coupling parameters
$c^{\rm pv}_{\Lambda_c\bar{D}^{(*)}\bar{K}^{*},\Lambda_b}$;
(ii) $\Sigma^{(*)}_c\bar{D}^{(*)}\to (J/\psi\, p)_{d {\rm -wave}}$ transitions
for which the HQSS gives an independent coupling.
In the $M_{J/\psi p}$ distribution of 
$\Lambda_b^0\to J/\psi pK^-$,
DT mechanisms from these unconsidered vertices would 
show singular behaviors similar to the considered DT mechanisms.
Thus, in the present analysis, we freely adjust the coupling parameters,
and effectively absorb these redundant mechanisms into 
the parameters of the considered mechanisms.
More detailed data are necessary to resolve the redundancy, thereby
determining the coupling parameters separately.
At this stage, we can examine how the HQSS constrains the fitting parameters.
\\

\begin{center}
 {2. $Y_c\bar D^{(*)}\to J/\psi p$ 
amplitudes} 
\end{center}

We describe a single-channel $Y_c\bar D^{(*)}$ $s$-wave
scattering (isospin 1/2) with an 
interaction potential as follows~\cite{d-decay}:
\begin{eqnarray}
v_\alpha (p',p) &=& 
\bigg(T_{Y_c}t_{Y_c}^\prime {1\over 2} t_{\bar{D}^{(*)}}^\prime\bigg| {1\over 2} {1\over 2} \bigg)
\bigg(T_{Y_c}t_{Y_c} {1\over 2} t_{\bar{D}^{(*)}}\bigg| {1\over 2} {1\over 2} \bigg)\nonumber\\
&&\times f^0_\alpha(p') h_\alpha\; f^0_\alpha(p) ,
\label{eq:cont-ptl}
\end{eqnarray}
where a label $\alpha$ specifies one of $Y_c\bar D^{(*)}$
and its $J^P$;
the isospin of $Y_c$ and its $z$-component are denoted by $T_{Y_c}$ 
and $t_{Y_c}^{(\prime)}$, respectively.
$h_\alpha$ is a coupling constant.
The form factor $f^0_\alpha$ has been defined in Eq.~(\ref{eq:ff2}).
With the above interaction potential, 
the elastic $Y_c\bar D^{(*)}$ scattering amplitude
is given as follows:
\begin{eqnarray}
t_\alpha (p',p; E) &=&
\bigg(T_{Y_c}t_{Y_c}^\prime {1\over 2} t_{\bar{D}^{(*)}}^\prime\bigg| {1\over 2} {1\over 2} \bigg)
\bigg(T_{Y_c}t_{Y_c} {1\over 2} t_{\bar{D}^{(*)}}\bigg| {1\over 2} {1\over 2} \bigg)\nonumber\\
&&\times
f^0_\alpha(p'){ h_\alpha \over
1 - h_\alpha \sigma_\alpha(E)}f^0_\alpha(p) ,
\label{eq:pw-2body-cont}
\end{eqnarray}
with
\begin{eqnarray}
\sigma_\alpha(E) &=&\!\sum_{\rm charge}
\bigg(T_{Y_c}t_{Y_c} {1\over 2} t_{\bar{D}^{(*)}}\bigg| {1\over 2}{1\over 2} \bigg)^2
\nonumber\\
&&\times \!
 \int\! dq q^2 
{ \left[f^0_\alpha(q)\right]^2 
\over E-E_{Y_c}(q)-E_{\bar{D}^{(*)}}(q) 
+{i\over 2}\Gamma_{Y_c} } ,
\label{eq:sigma}
\end{eqnarray}
where the summation runs over 
$\Sigma_c^{(*)+} \bar{D}^{(*)0}$ and 
$\Sigma_c^{(*)++} {D}^{(*)-}$ for $Y_c=\Sigma_c^{(*)}$;
the mass splitting between the different charge states are taken care of;
no summation for $Y_c=\Lambda_c^{(*,**)}$.
The tiny $\Gamma_{D^{*}}$ has been neglected
in Eq.~(\ref{eq:sigma}).
Assuming a perturbative
$Y_c\bar D^{(*)}\to J/\psi p$ interaction, 
we then obtain
\begin{eqnarray}
&&c_{\psi p, \alpha}^{J_\alpha^{P_\alpha}}
\bigg(T_{Y_c}t_{Y_c} {1\over 2} t_{\bar{D}^{(*)}}\bigg| {1\over 2} {1\over 2} \bigg)
f^{L_\alpha}_{\psi p}(p') f^0_\alpha(p) [1 - h_\alpha \sigma_\alpha(E)]^{-1},
\nonumber \\
\end{eqnarray}
for the $Y_c\bar D^{(*)}(J_\alpha^{P_\alpha})\to J/\psi p$ transition amplitude.
Now the DT amplitudes of our full model are obtained by multiplying
$[1 - h_\alpha \sigma_\alpha(E)]^{-1}$ with $E=M_{J/\psi p}$ to 
Eqs.~(5) and (\ref{eq:DT2})-(\ref{eq:DT3}).
The DT amplitudes without this modification are used in the simplified model.
Similarly, 
the one-loop amplitudes of the full model are obtained by 
multiplying
$[1 - h_\alpha \sigma_\alpha(E)]^{-1}$ 
to Eqs.~(8) and (\ref{eq:1L2})-(\ref{eq:1L3}).
The one-loop amplitudes without this modification are used in the
simplified model.
\\

\newpage
\begin{center}
 {3. Leading and lower-order singularities of double triangle diagrams}
\end{center}

The DT amplitudes of Eqs.~(\ref{eq:DT2})-(\ref{eq:DT3})
have lower-order singularities
in the zero-width limit of unstable particles in the loops.
Furthermore, some of the DT amplitudes have leading singularities, 
considering that the $K^*$ mass can have a range 
approximately as wide as its width. 
According to the Coleman-Norton theorem~\cite{coleman},
the DT leading singularity occurs only if the loop momenta hit a special
kinematical point where:
(i) $E=E_1=E_2=E_3$ (on-shell condition) with
$E_1\equiv E_{\bar{K}^*}+E_{\Lambda_c}+E_{\bar{D}^{(*)}}$,
$E_2\equiv E_{\bar{K}}+E_\pi+E_{\Lambda_c}+E_{\bar{D}^{(*)}}$, and 
$E_3\equiv E_{\bar{K}}+E_{\Sigma^{(*)}_c}+E_{\bar{D}^{(*)}}$;
(ii) the internal momenta are all collinear in the center-of-mass frame;
(iii) classically allowed kinematics
[$\hat{p}_{\bar{K}^*}\cdot\hat{p}_{\pi}=-1$,
$\hat{p}_{\bar{D}^{(*)}}\cdot\hat{p}_{\Sigma_c^{(*)}}=1$,
$v_\pi(\equiv |\bm{p}_\pi|/E_\pi)\ge v_{\Lambda_c^+}$,
$v_{\bar{D}^{(*)}}\ge v_{\Lambda_c^+}$, and
$v_{{\Sigma_c}^{(*)}}\ge v_{\bar{D}^{(*)}}$
for the case of $\hat{p}_{\Lambda_c}\cdot\hat{p}_{\bar{D}^{(*)}}=1$].
We examine realistic cases where the above conditions (ii) and (iii) are
satisfied but (i) is partly satisfied:
(i') $E=E_2=E_3\ne E_1$ and $|E-E_1|$ is minimum.
In these cases, the DT amplitudes have, at least, the lower-order singularity.

In Table~\ref{tab:DT2},
we present a set of particle momenta (center-of-mass frame)
from Eqs.~(\ref{eq:DT2})-(\ref{eq:DT3})
that satisfies the above conditions (i'), (ii) and (iii).
%
%
For this calculation, 
the $K^-$ momentum is set along the positive axis, 
and charge dependent masses are averaged.
Since $|E_1-E|\ltap \Gamma_{K^*}$
indicates the leading singularity,
$A^{\rm DT}_{\Sigma_c\bar{D}}$ and $A^{\rm DT}_{\Sigma_c\bar{D}^*}$  
($A^{\rm DT}_{\Sigma^*_c\bar{D}}$ and $A^{\rm DT}_{\Sigma^*_c\bar{D}^*}$)
have the leading (lower-order) singularities.
This fact 
and also a smaller effect from
$\Gamma_{\Sigma_c}<\Gamma_{\Sigma^*_c}$
($\Gamma_{\Sigma_c}\sim 2$~MeV,
$\Gamma_{\Sigma^*_c}\sim 15$~MeV)
 makes 
$A^{\rm DT}_{\Sigma_c\bar{D}}$ and $A^{\rm DT}_{\Sigma_c\bar{D}^*}$  
more singular than 
$A^{\rm DT}_{\Sigma^*_c\bar{D}}$ and $A^{\rm DT}_{\Sigma^*_c\bar{D}^*}$
near the 
$\Sigma^{(*)}_c\bar{D}^{(*)}$ thresholds,
as confirmed in Fig.~2.

\begin{table}[b]
\renewcommand{\arraystretch}{1.6}
\tabcolsep=1.5mm
\caption{\label{tab:DT2}
Particle momenta (center-of-mass frame) that 
satisfy the conditions (i'), (ii), and (iii) described in the text.
All momenta in the table are collinear.
$E_1\equiv E_{\bar{K}^*}+E_{\Lambda_c}+E_{\bar{D}^{(*)}}$.
The first column specifies the DT amplitude from 
Eqs.~(\ref{eq:DT2})-(\ref{eq:DT3}).
The unit of the entries is MeV.
}
\begin{tabular}[b]{c|ccccccc}
 & $p_{\bar K}$ & $p_{\bar K^*}$ & $p_\pi$ & $p_{\Lambda^+_c}$ & $p_{\bar{D}^{(*)}}$ &
			 $p_{\Sigma_c^{(*)}}$ & $E_1-E$ \\\hline
$A^{\rm DT}_{\Sigma_c\bar{D}}$    &1061 & 926 & $-135$ & $-471$ & $-455$ & $-607$ & $-76$ \\
$A^{\rm DT}_{\Sigma^*_c\bar{D}}$  &1006 & 771 & $-234$ & $-346$ & $-426$ & $-580$ & $-211$\\
$A^{\rm DT}_{\Sigma_c\bar{D}^*}$  & 937 & 807 & $-131$ & $-412$ & $-395$ & $-543$ & $-45$ \\
$A^{\rm DT}_{\Sigma^*_c\bar{D}^*}$& 879 & 654 & $-225$ & $-266$ & $-388$ & $-491$ & $-164$ \\
\end{tabular}
\end{table}


\begin{thebibliography}{}

\bibitem{qm}
S. Godfrey and N. Isgur,
Mesons in a relativized quark model with chromodynamics,
Phys. Rev. D {\bf 32}, 189 (1985).

 \bibitem{review_chen}
 H.-X. Chen, W. Chen, X. Liu, and S.-L. Zhu,
The hidden-charm pentaquark and tetraquark states,
Phys. Rep. {\bf 639}, 1 (2016).

 \bibitem{review_hosaka}
 A. Hosaka, T. Iijima, K. Miyabayashi, Y. Sakai, and S. Yasui,
Exotic hadrons with heavy flavors: $X$, $Y$, $Z$, and related states,
PTEP {\bf 2016}, 062C01 (2016).	 
	 
 \bibitem{review_lebed}
R.F. Lebed, R.E. Mitchell, and E.S. Swanson,
Heavy-Quark QCD Exotica,
Prog. Part. Nucl. Phys. {\bf 93}, 143 (2017).

 \bibitem{review_esposito}
A. Esposito, A. Pilloni, and A.D. Polosa,
 Multiquark Resonances,
Phys. Rept. {\bf 668}, 1 (2017).
	 
 \bibitem{review_ali}
A. Ali, J.S. Lange, and S. Stone, 
Exotics: Heavy Pentaquarks and Tetraquarks,
Prog. Part. Nucl. Phys. {\bf 97}, 123 (2017).

 \bibitem{review_guo}
F.-K. Guo, C. Hanhart, U.-G. Mei{\ss}ner, Q. Wang, Q. Zhao, and B.-S. Zou,
Hadronic molecules,
Rev. Mod. Phys. {\bf 90}, 015004 (2018).

 \bibitem{review_olsen}
S.L. Olsen, T. Skwarnicki, and D. Zieminska,
 Nonstandard heavy mesons and baryons: Experimental evidence,
	 Rev. Mod. Phys. {\bf 90}, 015003 (2018).

 \bibitem{review_Brambilla}
N. Brambilla, S. Eidelman, C. Hanhart, A. Nefediev, C.-P. Shen,
	 C.E. Thomas, A. Vairo, and C.-Z. Yuan,
The $XYZ$ states: Experimental and theoretical status and perspectives,
Phys. Rept. {\bf 873}, 1 (2020).

\bibitem{lhcb_pc}
R. Aaij et al. (LHCb Collaboration),
Observation of a Narrow Pentaquark State, $P_c(4312)^+$, 
and of the Two-Peak Structure of the $P_c(4450)^+$,
Phys. Rev. Lett. {\bf 122}, 222001 (2019).

\bibitem{lhcb_pc_old}
R. Aaij et al. (LHCb Collaboration),
Observation of $J/\psi p$ resonances consistent with pentaquark states in
$\Lambda_b^0\to J/\psi pK^-$ decays,
Phys. Rev. Lett. {\bf 115}, 072001 (2015).

\bibitem{pdg}
P.A. Zyla et al. (Particle Data Group), 
The Review of Particle Physics,
Prog. Theor. Exp. Phys. {\bf 2020}, 083C01 (2020).




\bibitem{pc_beihang}
M.-Z. Liu, Y.-W. Pan, F.-Z. Peng, M.S. S\'anchez, L.-S. Geng, A. Hosaka,
	and M.P. Valderrama,
Emergence of a complete heavy-quark spin symmetry multiplet: seven
	molecular pentaquarks in light of the latest LHCb analysis,
Phys. Rev. Lett. {\bf 122}, 242001 (2019).

\bibitem{pc_valencia}
C.-W. Xiao, J. Nieves, and E. Oset,
Heavy quark spin symmetric molecular states from 
$\bar{D}^{(*)}\Sigma_c^{(*)}$
and other coupled channels in the light of the recent LHCb pentaquarks,
Phys. Rev. D {\bf 100}, 014021 (2019).

\bibitem{pc_beihang2}
C.-J. Xiao, Y. Huang, Y.-B. Dong, L.-S. Geng, and D.-Y. Chen,
Exploring the molecular scenario of 
$P_c(4312)$, $P_c(4440)$, and $P_c(4457)$,
Phys. Rev. D {\bf 100}, 014022 (2019).

\bibitem{pc_hebei}
Z.-H. Guo and J.A. Oller,
Anatomy of the newly observed hidden-charm pentaquark states: 
$P_c(4312)$, $P_c(4440)$, and $P_c(4457)$,
 Phys. Lett. B {\bf 793}, 144 (2019).

\bibitem{pc_nanjin}
J. He,
Study of $P_c(4457)$, $P_c(4440)$, and $P_c(4312)$
in a quasipotential Bethe-Salpeter equation approach,
Eur. Phys. J. C {\bf 79}, 393 (2019).

\bibitem{pc_itp}
F.-K. Guo, H.-J. Jing, U.-G. Mei{\ss}ner, and S. Sakai,
Isospin breaking decays as a diagnosis of the hadronic molecular structure of the $P_c(4457)$,
Phys. Rev. D {\bf 99}, 091501(R) (2019).

\bibitem{pc_chen}
H.-X. Chen, W. Chen, and S.-L. Zhu,
Possible interpretations of the 
$P_c(4312)$, $P_c(4440)$, and $P_c(4457)$,
Phys. Rev. D {\bf 100}, 051501(R) (2019).

\bibitem{pc_lanzhou}
R. Chen, Z.-F. Sun, X. Liu, and S.-L. Zhu,
Strong LHCb evidence supporting the existence of the hidden-charm molecular pentaquarks,
Phys. Rev. D {\bf 100}, 011502(R) (2019).

\bibitem{pc_wang}
G.-J. Wang, L.-Y. Xiao, R. Chen, X.-H. Liu, X. Liu, and S.-L. Zhu,
Probing hidden-charm decay properties of $P_c$ states in a molecular scenario,
Phys. Rev. D {\bf 102}, 036012 (2020).

\bibitem{pc_gutsche}
T. Gutsche and V.E. Lyubovitskij,
Structure and decays of hidden heavy pentaquarks,
Phys. Rev. D {\bf 100}, 094031 (2019).

\bibitem{pc_peking}
B. Wang, L. Meng, and S.-L. Zhu,
Hidden-charm and hidden-bottom molecular pentaquarks in chiral effective field theory,
JHEP {\bf 11}, 108 (2019).

\bibitem{pc_nanjing2}
J. He and D.-Y. Chen,
Molecular states from $\Sigma^{(*)}_c\bar{D}^{(*)}$-$\Lambda_c\bar{D}^{(*)}$ interaction,
Eur. Phys. J. C {\bf 79}, 887 (2019).

\bibitem{pc_lin}
Y.-H. Lin and B.-S. Zou,
Strong decays of the latest LHCb pentaquark candidates in hadronic molecule pictures,
Phys. Rev. D {\bf 100}, 056005 (2019).

\bibitem{pc_burns}
T.J. Burns and E.S. Swanson,
Molecular interpretation of the $P_c(4440)$ and $P_c(4457)$ states,
Phys. Rev. D {\bf 100}, 114033 (2019).

\bibitem{pc_xu}
Y.-J. Xu, C.-Y. Cui, Y.-L. Liu, and M.-Q. Huang,
Partial decay widths of $P_c(4312)$ as a $\bar{D}\Sigma_{c}$ molecular state,
Phys. Rev. D {\bf 102}, 034028 (2020).

\bibitem{pc_yamaguchi}
Y. Yamaguchi, H. Garc\'ia-Tecocoatzi, A. Giachino, A. Hosaka, and E. Santopinto,
$P_c$ pentaquarks with chiral tensor and quark dynamics,
Phys. Rev. D {\bf 101}, 091502(R) (2020).

\bibitem{pc_sakai}
S. Sakai, H.-J. Jing, and F.-K. Guo,
Decays of $P_c$ into $J/\psi N$ and $\eta_c N$ with heavy quark spin symmetry,
Phys. Rev. D {\bf 100}, 074007 (2019).

\bibitem{pc_voloshin}
M.B. Voloshin,
Some decay properties of hidden-charm pentaquarks as baryon-meson molecules,
Phys. Rev. D {\bf 100}, 034020 (2019).

\bibitem{pc_wu}
Q. Wu and D.-Y. Chen,
Production of $P_c$ states from $\Lambda_b$ decay,
Phys. Rev. D {\bf 100}, 114002 (2019).

\bibitem{pc_jrzhang}
J.-R. Zhang,
Exploring a $\Sigma_{c}\bar{D}$ state: with focus on $P_c(4312)^+$,
Eur. Phys. J. C {\bf 79}, 1001 (2019).

\bibitem{pc_hxu}
H. Xu, Q. Li, C.-H. Chang, and G.-L. Wang,
Recently observed $P_c$ as molecular states and possible mixture of
	$P_c(4457)$,
Phys. Rev. D {\bf 101}, 054037 (2020).

\bibitem{pc_du}
M.-L. Du, V. Baru, F.-K. Guo, C. Hanhart, U.-G. Mei{\ss}ner, J.A. Oller, and Q. Wang,
Interpretation of the LHCb $P_c$ states as hadronic molecules and hints of a narrow $P_c(4380)$,
Phys. Rev. Lett. {\bf 124}, 072001 (2020).

\bibitem{pc_du2}
M.-L. Du, V. Baru, F.-K. Guo, C. Hanhart, U.-G. Mei{\ss}ner, J.A. Oller, and Q. Wang,
Revisiting the nature of the $P_c$ pentaquarks,
arXiv:2102.07159 [hep-ph].

\bibitem{pc_xiao}
C.-W. Xiao, J.-X. Lu, J.-J. Wu, and L.-S. Geng,
How to reveal the nature of three or more pentaquark states,
Phys. Rev. D {\bf 102}, 056018 (2020).


\bibitem{pc_ali}
A. Ali and A.Ya. Parkhomenko,
Interpretation of the narrow $J/\psi p$ peaks in
$\Lambda_b\to J/\psi pK^-$ decay in the compact diquark model,
Phys. Lett. B {\bf 793}, 365 (2019).

\bibitem{pc_pimikov}
A. Pimikov, H.-J. Lee, and P. Zhang,
Hidden charm pentaquarks with color-octet substructure in QCD Sum Rules,
Phys. Rev. D {\bf 101}, 014002 (2020).

\bibitem{pc_zgwang}
Z.-G. Wang,
Analysis of the $P_c(4312)$, $P_c(4440)$, $P_c(4457)$ and related
	hidden-charm pentaquark states with QCD sum rules,
Int. J. Mod. Phys. A {\bf 35}, 2050003 (2020).

\bibitem{pc_rzhu}
R. Zhu, X. Liu, H. Huang, and C.-F. Qiao,
Analyzing doubly heavy tetra- and penta-quark states by variational method,
Phys. Lett. B {\bf 797}, 134869 (2019).

\bibitem{pc_xzweng}
X.-Z. Weng, X.-L. Chen, W.-Z. Deng, and S.-L. Zhu,
Hidden-charm pentaquarks and $P_c$ states,
Phys. Rev. D {\bf 100}, 016014 (2019).

\bibitem{pc_bari}
F. Giannuzzi,
Heavy pentaquark spectroscopy in the diquark model,
Phys. Rev. D {\bf 99}, 094006 (2019).

\bibitem{pc_stancu}
F. Stancu,
Spectrum of the $uudc\bar{c}$ hidden charm pentaquark with an SU(4)
	flavor-spin hyperfine interaction,
Eur. Phys. J. C {\bf 79}, 957 (2019).

\bibitem{pc_ydong}
Y. Dong, P. Shen, F. Huang, and Z. Zhang,
Selected strong decays of pentaquark State $P_c(4312)$ in a chiral constituent quark model,
Eur. Phys. J. C {\bf 80}, 341 (2020).

\bibitem{pc_hadrochamonium}
M.I. Eides, V.Yu Petrov, and M.V. Polyakov,
New LHCb pentaquarks as hadrocharmonium states,
Mod. Phys. Lett. A {\bf 35}, 2050151 (2020).


\bibitem{jpac}
C. Fern\'andez-Ram\'irez, A. Pilloni, M. Albaladejo, A. Jackura,
	V. Mathieu, M. Mikhasenko, J.A. Silva-Castro, and A.P. Szczepaniak,
Interpretation of the LHCb $P_c(4312)^+$ Signal,
Phys. Rev. Lett. {\bf 123}, 092001 (2019).


\bibitem{pc_kuang}
S.-Q. Kuang, L.-Y. Dai, X.-W. Kang, and D.-L. Yao, 
Pole analysis on the hadron spectroscopy of $\Lambda_b^0\to J/\psi pK^-$,
Eur. Phys. J. C {\bf 80}, 433 (2020).

\bibitem{photo_qwang}
Q. Wang, X.-H. Liu, and Q. Zhao,
Photoproduction of hidden charm pentaquark states $P_c^+(4380)$
and $P_c^+(4450)$,
 Phys. Rev. D {\bf 92}, 034022 (2015).

\bibitem{photo_kubarovsky}
V. Kubarovsky and M.B. Voloshin,
Formation of hidden-charm penta quarks in photon-nucleon collisions,
Phys. Rev. D {\bf 92}, 031502(R) (2015).

\bibitem{photo_Karliner}
M. Karliner and J.L. Rosner,
Photoproduction of Exotic Baryon Resonances,
Phys. Lett. B {\bf 752}, 329 (2016).

\bibitem{photo_hiller}
Studying the $P_c(4450)$ resonance in $J/\psi$ photoproduction off protons,
A.N. Hiller Blin, C. Fern\'andez-Ram\'irez, A. Jackura, V. Mathieu,
	V.I. Mokeev, A. Pilloni, and A.P. Szczepaniak,
Phys. Rev. D {\bf 94}, 034002 (2016).

\bibitem{photo_xywang}
X.-Y. Wang, X.-R. Chen, and J. He,
Possibility to study pentaquark states
$P_c(4312)$, $P_c(4440)$, and $P_c(4457)$ in $\gamma p\rightarrow J/\psi p$ reaction,
Phys. Rev. D {\bf 99}, 114007 (2019).

\bibitem{photo_wu}
J.-J. Wu, T.-S.H. Lee, and B.-S. Zou,
Nucleon resonances with hidden charm in $\gamma p$ reactions,
Phys. Rev. C {\bf 100}, 035206 (2019).

\bibitem{photo_cao}
X. Cao and J.-P. Dai,
Confronting pentaquark photoproduction with new LHCb observations,
Phys. Rev. D {\bf 100}, 054033 (2019).

\bibitem{gluex}
A. Ali et al. (GlueX Collaboration), 
First measurement of near-threshold $J/\psi$ exclusive photoproduction off the proton,
Phys. Rev. Lett. {\bf 123}, 072001 (2019).

\bibitem{ts_review}
F.-K. Guo, X.-H. Liu, and S. Sakai,
Threshold cusps and triangle singularities in hadronic reactions,
Prog. Part. Nucl. Phys. {\bf 112}, 103757 (2020).

\bibitem{ts1}
F.-K. Guo, U.-G. Mei{\ss}ner, W. Wang, and Z. Yang,
How to reveal the exotic nature of the $P_c(4450)$,
Phys. Rev. D {\bf 92}, 071502(R) (2015).

\bibitem{ts2}
X.-H. Liu, Q. Wang, and Q. Zhao,
Understanding the newly observed heavy pentaquark candidates,
Phys. Lett. B {\bf 757}, 231 (2016).

\bibitem{s-matrix}
R. J. Eden, P. V. Landshoff, D. I. Olive and J. C. Polkinghorne,
The Analytic S-Matrix,
(Cambridge University Press, Cambridge, England, 1966).

\bibitem{Pc_lhcb2}
R. Aaij et al. (LHCb Collaboration),
Evidence for exotic hadron contributions to 
$\Lambda_b^0 \to J/\psi p \pi^-$ decays,
Phys. Rev. Lett. {\bf 117}, 082003 (2016).

\bibitem{sxn_x}
S.X. Nakamura, 
Triangle singularity appearing as an $X(3872)$-like peak in 
$B\to (J/\psi\pi^+\pi^-) K\pi$,
Phys. Rev. D {\bf 102}, 074004 (2020).

\bibitem{lqcd_jpsi_p}	
 U. Skerbis and S. Prelovsek, 
Nucleon-$J/\psi$ and nucleon-$\eta_c$ scattering in $P_c$ pentaquark channels from LQCD,
Phys. Rev. D {\bf 99}, 094505 (2019).

\bibitem{d-decay}
S.X. Nakamura,
Coupled-channel analysis of $D^+\to K^-\pi^+\pi^+$ decay,
Phys. Rev. D {\bf 93}, 014005 (2016).

\bibitem{coleman}
S. Coleman and R.E. Norton,
Singularities in the physical region ,
Nuovo Cimento {\bf 38}, 438 (1965).

\bibitem{landau}
L.D. Landau, 
On analytic properties of vertex parts in quantum field theory,
Nucl. Phys. {\bf 13}, 181 (1959).

\bibitem{xkdong}
X.-K. Dong, F.-K. Guo, and B.-S. Zou,
Explaining the Many Threshold Structures in the Heavy-Quark Hadron Spectrum,
Phys. Rev. Lett. {\bf 126}, 152001 (2021).

	
\bibitem{3pi}
H. Kamano, S.X. Nakamura, T.-S.H. Lee, and T. Sato,
Unitary coupled-channels model for three-mesons decays of heavy mesons,
Phys. Rev. D {\bf 84}, 114019 (2011).


\end{thebibliography}


\end{document}